\documentclass[%
 reprint,
 amsmath,amssymb,
 aps,
pre,
floatfix,
]{revtex4-1}

\usepackage{graphicx}
\usepackage{dcolumn}
\usepackage{bm}
\usepackage{subfigure}
\usepackage{hyperref}
\usepackage{wrapfig}
\usepackage{placeins}
\usepackage{natbib}

\begin{document}


\title{\textbf{ Ferroelectric domain formation in discotic liquid crystals : Monte Carlo study on the influence of boundary conditions}}


\author{Tushar Kanti Bose}
\email{tkb.tkbose@gmail.com}
\author{Jayashree Saha}
\email{jsphy@caluniv.ac.in}
\affiliation{Department of Physics, University of Calcutta, 92 Acharya Prafulla Chandra Road, Kolkata 700009, India}
 




\date{\today}

\begin{abstract}

The realization of a spontaneous macroscopic ferroelectric order in fluids of anisotropic mesogens is a topic of both fundamental and technological interest. Recently, we demonstrated that a system of dipolar achiral disklike ellipsoids can exhibit long-searched ferroelectric liquid crystalline phases of dipolar origin. In the present work, extensive off-lattice Monte Carlo simulations are used to investigate the phase behavior of the system under the influences of the electrostatic boundary conditions that restrict any global polarization. We find that the system develops strongly ferroelectric slablike domains periodically arranged in an antiferroelectric fashion. Exploring the phase behavior at different dipole strengths, we find existence of the ferroelectric nematic and ferroelectric columnar order inside the domains. For higher dipole strengths, a biaxial phase is also obtained with a similar periodic array of ferroelectric slabs of antiparallel polarizations. We have studied the depolarizing effects by using both the Ewald summation and the spherical cut-off techniques. We present and compare the results of the two different approaches of considering the depolarizing effects in this anisotropic system. It is explicitly shown that the domain size increases with the system size as a result of considering longer range of dipolar interactions. The system exhibits pronounced system size effects for stronger dipolar interactions. The results provide strong evidence to the novel understanding that the dipolar interactions are indeed sufficient to produce long range ferroelectric order in anisotropic fluids. 






\end{abstract}

\pacs{Valid PACS appear here}
\maketitle


\section{\label{sec:level1}Introduction }


The understanding and development of various possible polar order in liquid crystals is an important area of research in soft matter physics and chemistry \cite{b1,b2,b3,b4,b5}. The studies of ferroelectric and antiferroelectric fluids has attracted much attention in the last 25 years \cite{b1,b3,b4,b5}. The systems of chiral rodlike molecules and achiral bent core molecules are the conventional systems which exhibited ferroelectric and antiferroelectric liquid crystal phases. Chiral disk shaped molecules also exhibit ferroelectric columnar phases \cite{b8}. 
 Usually in the conventional ferroelectric smectic and columnar phases, the respective ferroelectric orderings are not primarily driven by dipole-dipole interactions. 
However, there is no fundamental reason to forbid a proper ferroelectric fluid with spontaneous ferroelectric order of dipolar origin \cite{b9,b10,b11,b12,b13,b14,b15,b16,b17,b21}. From computer simulation studies, it has been found that model spherical particles with a strong central dipole moment exhibit ferroelectric fluid phases in conducting surroundings \cite{b10,b11,b21,b22,b221}. In some studies, the dipolar spheres showed the formation of ferroelectric domains in the presence of a strong depolarizing field in insulating vacuum surroundings \cite{b21,b22}. The spontaneous ferroelectric order in the system of dipolar spheres is developed solely due to the dipolar interactions and the lack of orientational bias of the spherical particles. On the contrary, the situation is very different for aspherical particles having dipole moments. A spontaneous macroscopic ferroelectric order of dipolar origin is very rare in fluids of anisotropic molecules which are the natural candidates to form liquid crystal phases. It was also shown that as the length to breadth ratio decreases from unity, the tendency to form a ferroelectric nematic phase gradually decreases for discotic ellipsoidal particles with strong central dipoles \cite{b2222}. Recently, we have shown that the realization of a novel class of proper ferroelectric liquid crystal phases is possible in a system of achiral dipolar disklike Gay-Berne(GB) ellipsoids in conducting surroundings \cite{b23}. The model system of interest consists of attractive-repulsive GB oblate ellipsoids embedded with two parallel point dipoles positioned symmetrically on the equatorial plane of the ellipsoids. The system exhibited stable ferroelectric nematic and columnar fluids with strong overall polarization. The study demonstrated that the dipolar interactions are indeed sufficient to produce a class of novel ferroelectric fluids of essential interest in systems of disc shaped particles \cite{b23}. A system exhibiting a spontaneous macroscopic polarization is usually expected to be sensitive to the electrostatic boundary conditions. Thus, it becomes interesting enough to investigate the influences of the boundary conditions upon the novel ferroelectric liquid crystal phases of interest. Here, we study the phase behavior of the system under the influence of the boundary condition that restricts any global polarization. We study the phase behavior of the system using the simple spherical cutoff approach excluding the contribution of a polarizing reaction field \cite{b284} and also with the more conventional Ewald summation method \cite{b284} incorporating the effect of a strong depolarizing field. Both the methods are described in Sec. \ref{sec:level2}\\

Extensive work focused on testing the influence of different boundary conditions on the behavior of simple models of dipolar liquids, for example, the dipolar hard spheres and Stockmayer fluid, was started in the 1980s \cite{b28406,b28401,b28402,b28403,b28404,b28405}. Even recently computational studies of dipolar systems have been performed using different boundary conditions \cite{b333,b3336}. However, the present work is one of the firsts in testing the effects of boundary conditions on an anisotropic system exhibiting spontaneous polarization of dipolar origin. It must be emphasized that the attractive-repulsive GB ellipsoids can naturally form liquid crystal phases even without dipolar interactions whereas the spherically symmetric particles are unable to form orientationally ordered phases without dipolar interactions.\\

Now, it can be easily understood that the problem of determining the collective organization of a system of dipolar molecules is definitely non-trivial, since it is the result of a balance between the tendency of two dipoles to arrange anti-parallel and the formation of domains with a common dipole orientation that gives fascinating domain structures \cite{b22,b221,b2841}. 
In the present work, we systematically investigate the influence of the depolarizing boundary condition upon the existence of different ferroelectric phases of interest. The system results in the formation of slablike ferroelectric fluid domains periodically arranged in an antiferroelectric fashion. More importantly, at different state points and dipole strengths, we find existence of long range ferroelectric nematic, ferroelectric columnar and ferroelectric biaxial order within the domains. Systems of different size (N = 1500, 4000 and 8500) have been investigated via the simple spherical cutoff approach. It is explicitly shown that the size of the ferroelectric domains grow with the system size as an effect of considering longer range of dipolar interactions. In Sec. \ref{sec:level2} we detail the molecular model and pair interactions. The simulation methods are described in Sec. \ref{sec:level3}. In Sec. \ref{sec:level4} we describe the simulation results and Sec. \ref{sec:level5} concludes the paper. 

\section{\label{sec:level2}Model }
In this paper we present computer simulations of a system of uniaxial oblate GB ellipsoids of revolution where each ellipsoid is embedded with two axial off-center parallel point dipole moments. The dipoles are symmetrically placed on the equatorial plane of the ellipsoid, at equal distances from the center of the ellipsoid. In the present work, the dipoles are placed on the molecular x axis (perpendicular to the symmetry axis) of each GB molecule, separated by a distance \(d^{*}\equiv d/\sigma_{0}=0.5\) along the axis. The dipolar ellipsoids are interacting via a pair potential which is a sum of a modified form of the GB potential \cite{b29} and the electrostatic dipolar interactions. In the modified form for discotic liquid crystal \cite{b30}, the pair potential between two oblate ellipsoids \textit{i} and \textit{j} is given by 
\[ \displaystyle{ U_{ij}^{\textrm{GB}}(\mathbf{r}_{ij},\mathbf{\hat{u}}_{i},\mathbf{\hat{u}}_{j})=4\epsilon(\mathbf{\hat{r}}_{ij},\mathbf{\hat{u}}_{i},\mathbf{\hat{u}}_{j})(\rho_{ij}^{-12}-\rho_{ij}^{-6}) } \]
\(\displaystyle{\mbox{ where }\rho_{ij} = ( r_{ij} - \sigma( \mathbf{r}_{ij},\mathbf{\hat{u}}_{i},\mathbf{\hat{u}}_{j} ) + \sigma_{e} ) / \sigma_{e} \mbox{.}}\) Here unit vectors \(\mathbf{\hat{u}}_{i}\mbox{ and }\mathbf{\hat{u}}_{j}\) represent the orientations of the symmetry axes of the molecules, \(\mathbf{r}_{ij}=r_{ij}\mathbf{\hat{r}}_{ij}\) is the separation vector of length \(r_{ij}\) between the centers of mass of the ellipsoids and \(\sigma_{e}\) is the minimum separation between two ellipsoids in a face-to-face configuration determining the thickness of the ellipsoids. The anisotropic contact distance \(\sigma\) and the depth of pair interaction well \(\epsilon\) are dependent on four important parameters \(\kappa,\kappa ',\mu,\nu\), as defined in \cite{b30}. Here \(\kappa=\sigma_{e}/\sigma_{0}\) is the aspect ratio of the ellipsoids where \(\sigma_{0}\) is the minimum separation between two ellipsoids in a side-by-side configuration, \(\kappa '=\epsilon_{s}/\epsilon_{e}\) is ratio of interaction well depths in side-by-side and face-to-face configuration of the disc shaped ellipsoids. The other two parameters \(\mu\) and \(\nu\) control the well depth of the potential. \(\sigma_{0}\) and \(\epsilon_{0}\) define the length and energy scales respectively where \(\epsilon_{0}\) is the well depth in the cross configuration. The values used here to study the bulk phase behavior are \(\kappa=0.345,\kappa'=0.2, \mu=1, \nu=3\). The value of \(\kappa\) is obtained from the parametrization of the GB potential that mimics the interaction between two molecules of triphynylene \cite{b31} which is known to form the core of many discotic mesogens \cite{b32}. The other parameters were chosen from previous works on discotic liquid crystals which exhibited discotic nematic and hexagonal columnar phases \cite{b33,b282,b331}. The electrostatic interaction energy between two such dipolar ellipsoids is given by \(\displaystyle{ {U_{ij}^{\textrm{dd}}}=\sum_{\alpha, \beta=1}^{2}\frac{{\mu}^{2}}{ r_{\alpha \beta}^{3}}\left[(\bm{\hat{\mu}}_{i \alpha}\cdot{\bm{\hat{\mu}}}_{j \beta})-3(\bm{\hat{\mu}}_{i \alpha}\cdot\bm{\hat{r}}_{\alpha \beta})(\bm{\hat{\mu}}_{j \beta}\cdot\bm{\hat{r}}_{\alpha \beta}) \right]  }  \) , where \(\textbf{r}_{\alpha \beta} (=\textbf{r}_{j \beta}-\textbf{r}_{i \alpha}) \) is the vector joining the two point dipoles \( \boldsymbol{\mu}_{i\alpha} \) and \( \boldsymbol{\mu}_{j\beta} \) on the molecules \textit{i} and \textit{j} at the positions \(\textbf{r}_{i \alpha} = \textbf{r}_{i}\pm \frac{d}{2}\bm{\hat{x}_{i}}\) and \(\textbf{r}_{j \beta} = \textbf{r}_{j}\pm \frac{d}{2}\bm{\hat{x}}_{j}\). Then the total interaction energy between two dipolar molecule is given by \( U_{ij}^{\textrm{total}}= U_{ij}^{\textrm{GB}}+U_{ij}^{\textrm{dd}} \). Here we have used different values of the reduced dipole moments \( \mu^{*} \equiv \sqrt{\mu^{2}/\epsilon_{0}\sigma_{0}^{3}} = 0.2,0.4,0.6,0.7 \mbox{ and }0.9 \) to investigate the effect of dipole strength on the phase behavior of the system of interest. The dipole moments \( \mu^{*} = 0.2 \mbox{ and } 0.9 \), for a molecular diameter of \(\sigma_{0}\approx 10\AA\) and an energy term \(\epsilon_{0}=5\times10^{-15}\mbox{ erg}\) corresponds to 0.45 D and 2 D respectively. 

The Ewald summation (ES) and reaction field (RF) methods are two highly popular methods to handle the long range dipolar interaction in condensed matter systems. In the ES method \cite{b284}, the central cubic simulation cell is surrounded by its replicas in all directions to form an infinitely large sphere of periodic images. This infinitely large sphere is itself embedded in a continuous medium of dielectric constant \(\epsilon_{s}\). The expression for the overall dipolar interaction energy of the system in the ES technique is given by\\

\(\displaystyle{  U_{ES} =  - \frac{1}{2} \sum_{i}  \sum_{j}  {\sum_{n}}^{'} (\boldsymbol{\mu}_{i}.\boldsymbol{\nabla})  (\boldsymbol{\mu}_{j}.\boldsymbol{\nabla})\frac{erfc(\alpha |\boldsymbol{r}_{ij}+\boldsymbol{n}L|)}{|\boldsymbol{r}_{ij}+\boldsymbol{n}L|}}\\ \hspace*{2.25cm} \displaystyle{+\frac{1}{2\pi V} \sum_{\boldsymbol{k}\neq 0} \frac{4 \pi^{2}}{k^{2}} F(\boldsymbol{k})F^{*}(\boldsymbol{k}) exp \left(\frac{-k^{2}}{4\alpha^{2}} \right)}\\ \hspace*{2.25cm} \displaystyle{-\sum_{i} \frac{2\alpha^{3}{\mu_{i}}^{2}}{3\sqrt{\pi}} + \frac{2\pi}{(2\epsilon_{s}+1)V} M^{2} }\)\\

where \( \displaystyle{ F(\boldsymbol{k}) = \sum_{i} (\boldsymbol{k}.\boldsymbol{\mu}_{i}) exp(i \boldsymbol{k}.\boldsymbol{r}_{i}) } \) and
      \(\displaystyle{   \boldsymbol{M} = \sum_{i} \boldsymbol{\mu}_{i}   }\).


where \(\boldsymbol{r}_{i}\) is the position vector of the dipole moment \(\boldsymbol{\mu}_{i}\). \(\boldsymbol{M}\) is the total dipole moment of the cubic central box of volume V=\(L^{3}\), erfc is the complementary error function and the sum on \textbf{n} is over lattice vectors. The prime in the sum over \textbf{n}=(\(n_{x},n_{y},n_{z}\)), \(n_{x},n_{y},n_{z}\) integers, indicates that \(i \neq j\) for \(|\boldsymbol{n}|=0\). The reciprocal space vectors are of the form \(\displaystyle{\boldsymbol{k}=\frac{2\pi}{L}\boldsymbol{n}}\). The convergence of the interaction terms in the ES depends on the parameter \(\alpha\) which is chosen such that the real space sum over lattice vectors can be restricted to the first term in the sum i.e. the term with \(|\boldsymbol{n}|=0\).  The last term in the above equation for \(U_{ES}\), accounts for the work done against the depolarizing field due to the surface charges induced on the spherical boundary. This term vanishes only for \(\epsilon_{s}\rightarrow\infty\)(conducting / tin-foil boundary condition) and makes its maximum contribution when \(\epsilon_{s}=1\) ( vacuum boundary condition). The contribution to this term from the pair interactions is given by \( \displaystyle{ \frac{4\pi}{(2\epsilon_{s}+1) L^{3}}\boldsymbol{\mu}_{i}.\boldsymbol{\mu}_{j}  }\). It is to be noted that the term does not decay with separation and it is clear from \(\boldsymbol{\mu}_{i}.\boldsymbol{\mu}_{j}\) dependence that this term will favor antiparallel rather than parallel dipolar orientations. The choice of \(\epsilon_{s}=1\) effectively results in surrounding the large spherical sample of cubic replicas with a vacuum and is extreme for a ferroelectric medium. Therefore we can expect to observe a significant effect upon the ferroelectric order found in \cite{b23} when \(\epsilon_{s}\) is set equal to 1. The Ewald sum parameters used in the present work for a system of N = 500 dipolar ellipsoids are : ( \(|\textbf{n}_{max}|\) = 3, \(\textbf{n}^{2}_{max}=27\) containing 334 \(\bm{\vec{k}}\) vectors , \(\alpha = 15\) ).\\

In the reaction field method \cite{b284}, the dipolar pair interactions are exactly evaluated upto a cutoff radius \(R_{C}\) and the dipoles outside cutoff sphere are considered as a continuous medium with dielectric constant \(\epsilon_{s}\). Polarization of the continuous medium in response to the polarization from the total dipole moment \textbf{M} within cutoff sphere produces an electric field \(\boldsymbol{R}_{i}\) that acts back on the i th dipole, located at the center of the cavity. The reaction field is expressed as \(\displaystyle{\boldsymbol{R}_{i}=\frac{2(\epsilon_{s}-1)}{(2\epsilon_{s}+1){R_{C}}^{3}}\boldsymbol{M} }\). The size of the reaction field along any molecule is proportional to the moment of the cavity surrounding that molecule. The contribution to energy from reaction field is \(-\frac{1}{2}\boldsymbol{\mu}_{i}.\boldsymbol{R}_{i}\), which is a negative contribution if \(\boldsymbol{\mu}_{i}\) is parallel to \(\boldsymbol{M}\). Therefore, the reaction field tends to polarize the system when a ferroelectric phase is formed, by reducing the potential energy of the system in favor of global ferroelectric order.\\

The value of \(\epsilon_{s}\) strongly influences the behavior of a ferroelectric system. If \(\epsilon_{s}\rightarrow\infty\)(conducting or tin-foil boundary condition), the contribution to energy from reaction field strongly favors a polarization. As mentioned before, the systems of dipolar spheres exhibited global ferroelectric order under the influence of the conducting boundary conditions in both the ES \cite{b10,b11,b21,b22,b221} and RF methods \cite{b332,b3321}. The present system of dipolar ellipsoids was also found to generate global ferroelectric order in our previous works \cite{b23,b4444} under the influence of the conducting boundary conditions. If \(\epsilon_{s}=1\)(vacuum or spherical cutoff boundary condition), \(\boldsymbol{R}_{i}=0\), so only pairwise interactions within cavity are considered. This boundary condition is used to simulate the behavior of a macroscopic sample placed in vacuum and acts against the formation of global polarization. If \(\epsilon_{s}=1\), the system tends to form local structures while keeping the total polarization zero. The RF method with \(\epsilon_{s}=1\) has been satisfactorily used in studies of the orientationally ordered systems of dipolar spheres \cite{b332,b333,b3336} and model systems of water \cite{b334,b335}.\\

 It should be noted that in the case of ES a depolarizing field appears for \(\epsilon_{s}=1\), which is absent for \(\epsilon_{s}\rightarrow\infty\). In the RF method, a polarizing field appears for \(\epsilon_{s}\rightarrow\infty\), absent when \(\epsilon_{s}=1\), with the same effect. So, we expect to observe qualitatively similar phase behavior using both the methods.\\

We have used three different system size : N = 1500 (\(R_{C}=3\sigma_{0}\)), N = 4000 (\(R_{C}=5\sigma_{0}\)) and N = 8500 (\(R_{C}=7\sigma_{0}\)) when investigating the phase behaviour of the system using spherical cutoff approach ( \(\epsilon_{s}=1\), \(\boldsymbol{R}_{i}=0\) ) .



\begin{figure*}[!]
\centering
\subfigure[\label{fig:p7a}]{\includegraphics[scale=.75]{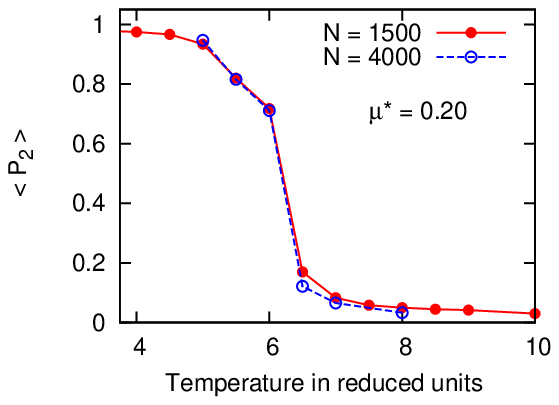}}\hspace{-0.25cm}
\subfigure[\label{fig:p7b}]{\includegraphics[scale=.75]{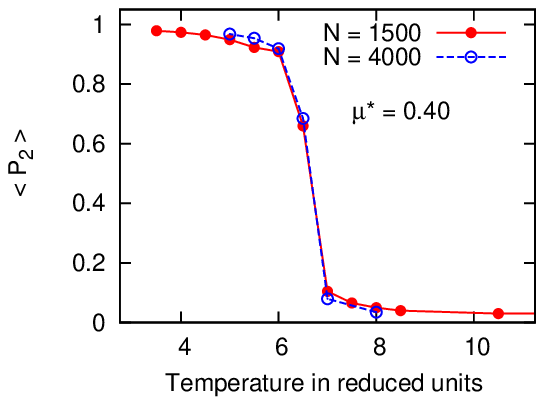}}\hspace{-0.25cm}
\subfigure[\label{fig:p7c}]{\includegraphics[scale=.75]{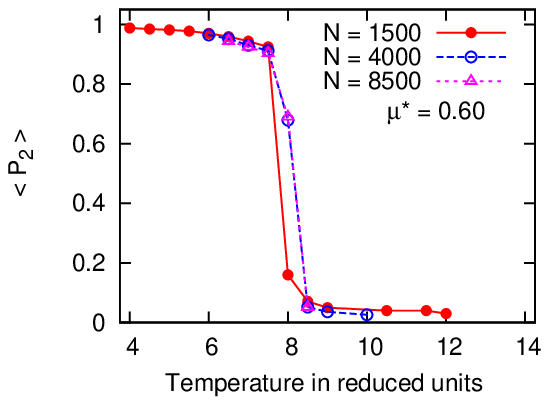}}\hspace{-0.25cm}
\subfigure[\label{fig:p7d}]{\includegraphics[scale=.75]{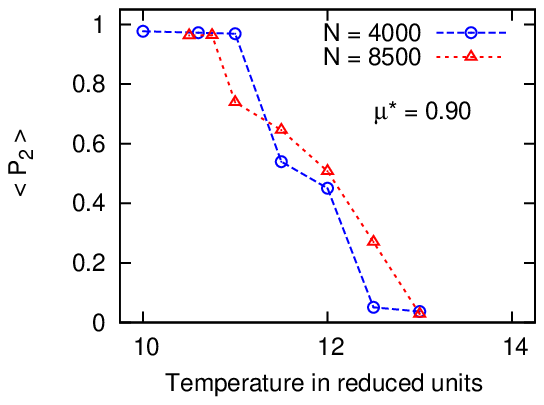}}

\caption{\label{fig:p7}(color online).(a) Evolution of the average second-rank order parameter \(\langle P_{2}\rangle\) against reduced temperature \(T^{*}\) at constant pressure \(P^{*}=100.0\) for \(\mu^{*}\)=0.20 and \textit{N} = 1500, 4000. (b) Evolution of \(\langle P_{2}\rangle\) against \(T^{*}\) at \(P^{*}=100.0\) for \(\mu^{*}\)=0.40 and \textit{N} = 1500, 4000. (c) Evolution of \(\langle P_{2}\rangle\) against \(T^{*}\) at \(P^{*}=100.0\) for \(\mu^{*}\)=0.60 and \textit{N} = 1500, 4000 and 8500. (d)  Evolution of \(\langle P_{2}\rangle\) against \(T^{*}\) at \(P^{*}=100.0\) for \(\mu^{*}\)=0.90 and \textit{N} = 4000 and 8500..
 }
\end{figure*}

\section{\label{sec:level3}Simulation Details }
We have performed Monte Carlo (MC) simulation studies in the isothermal-isobaric ( constant NPT ) ensemble with periodic boundary conditions imposed on the systems of dipolar ellipsoids. We have performed a cooling sequence of simulation runs along an isobar at fixed pressure \(P^{*}(\equiv P\sigma_{0}^{3}/\epsilon_{0})=100\). A MC simulation of the system of GB ellipsoids without dipoles yielded a discotic nematic and hexagonal columnar phases at the same pressure without any ferroelectric order \cite{b282}. We started the simulation from an well equilibrated isotropic liquid phase in a cubic box. We then reduced the temperature of the system sequentially to explore the phase behavior. At a given temperature, the final equilibrated configuration obtained from the previous higher temperature was used as the starting configuration. The system was subjected to long equilibrium runs at each state point [\(p^{*},T^{*}\)]. During a MC cycle, each particle was randomly displaced and reoriented following metropolis criteria where the reorientation moves were performed using Barker-Watts technique \cite{b284}. An attempt to change the volume of the cubic box was also performed in each MC cycle. The acceptance ratio of the roto-translational moves and volume were adjusted to 40\%. To overcome any possibility of locking in a metastable state, the particles were also allowed to attempt up-down flip moves exchanging particle tip with bottom with a 20\% frequency with respect to the roto-translational MC moves. We have also used an orthogonal box at some state points.
\section{\label{sec:level4}Results \& Discussions }

In order to characterize different phases of the system, various order parameters were computed. 
The average orientational order of the particles was monitored by the second-rank orientational order parameter \(P_{2}\) defined by the largest eigenvalue of the order tensor \(S_{\alpha\beta}=\frac{1}{N}\sum_{i=1}^{N}\frac{1}{2}(3u_{i\alpha}u_{j\beta}-\delta_{\alpha\beta})\), where \(\alpha,\beta=x,y,z\) are the indices referring to three components of the unit vector \(\hat{\textbf{u}}\) along the orientation of the particles and \(\delta_{\alpha\beta}\) is the Kronecker delta symbol. The value of \(P_{2}\) is close to zero in the isotropic phase and tends to 1 in the highly ordered phases. The global ferroelectric order was measured by calculating the average polarization per particle \(P_{1}\) defined by \(P_{1}=\frac{1}{N}\sum_{i=1}^{N}\) \( \hat{\mu_{i}}. \hat{{d}} \) where \(\hat{\textbf{d}}\) is the director of the system. \(P_{1}\) is unity in a perfectly ferroelectric phase and zero in an antiferroelectric phase and in the isotropic phase. \(P_{2}\) is therefore the indicator of global orientational order and \(P_{1}\) distinguishes between ferroelectric and antiferroelectric phases. We have also measured the biaxial order parameter  \( \langle R_{2,2}^{2}\rangle=\langle\frac{1}{2}(1+\cos^{2}\beta) \cos 2\alpha \cos 2\gamma -\cos\beta \sin2 \alpha \sin2 \gamma \rangle     \) as described in \cite{b34}, where \(\alpha,\beta,\gamma \) are the Euler angles giving the orientation of the molecular body set of axes w.r.t. the director set of axes. In a biaxial phase, the anisotropic molecules exhibit additional orientational order along a second macroscopic direction perpendicular to the primary director. It means that we can define a set of perpendicular macroscopic axes of preferential orientation (only two need to be defined as the third is then specified as perpendicular to the other two) in a biaxial phase. In the present system, the symmetry axes of the ellipsoids remain aligned in all the uniaxial and biaxial ferroelectric discotic phases but the molecular x axes ( axes along the separation between two dipoles on each ellipsoid) and molecular y axes ( axes perpendicular to both the molecular symmetry axis and molecular x axis) are significantly oriented only in a biaxial phase. The biaxial order parameter \(\langle R_{2,2}^{2} \rangle\) measures the degree of ordering of the molecular x and y axes in a plane perpendicular to the primary director \cite{b34}. For \(\langle P_{2} \rangle\) = 1, \(\langle R_{2,2}^{2} \rangle\) = 0 the system is perfectly uniaxial and for \(\langle P_{2} \rangle\) = 1, \(\langle R_{2,2}^{2} \rangle\) = 1 the system is perfectly biaxial. 
In order to verify the fluidity of the ferroelectric phases, we calculated the mean square displacement as follows : \( \langle R^{2} \rangle_{\tau} = \frac{1}{N} \sum_{i=1}^{N}[ \textbf{r}_{i}(\tau)-\textbf{r}_{i}(0)]^{2} \) , where \(\textbf{r}_{i}(\tau)\) is the position vector of the i th particle after completion of \(\tau\) MC cycles. In the fluid phases, the mean square displacement steadily increases with increasing \(\tau\) indicating fluid behavior. In contrast for solids \(\langle R^{2} \rangle_{\tau}\) becomes constant as \(\tau\) increases. For a proper structural analysis of the resultant ferroelectric phases, we calculated important distribution functions as required. We have measured the radial distribution function \(g(r)=\frac{1}{4\pi r^{2}\rho}\langle\delta(r-r_{ij})\rangle_{ij}\mbox{ ,}\)where the average is taken over all the molecular pairs. The columnar distribution function \(g_{c}(r_{\shortparallel}^{*})\) and the perpendicular distribution function \(g(r_{\perp}^{*})\) for the disklike ellipsoids were calculated following \cite{b30}.\\ 
\begin{figure*}[p]
\centering
\subfigure[\label{fig:p1a } ]{\includegraphics[scale=0.75]{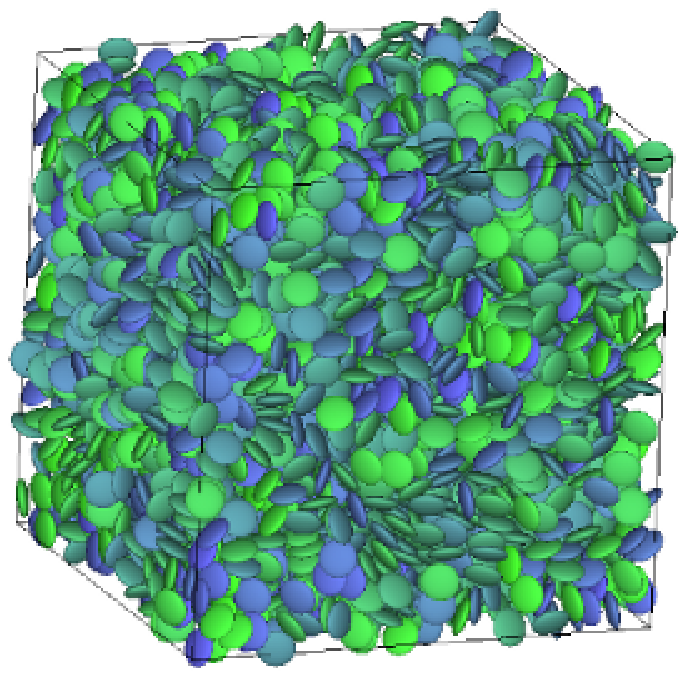}}\hspace{0.2cm}
\subfigure[\label{fig:p1b }]{\label{nematic}\includegraphics[scale=0.65]{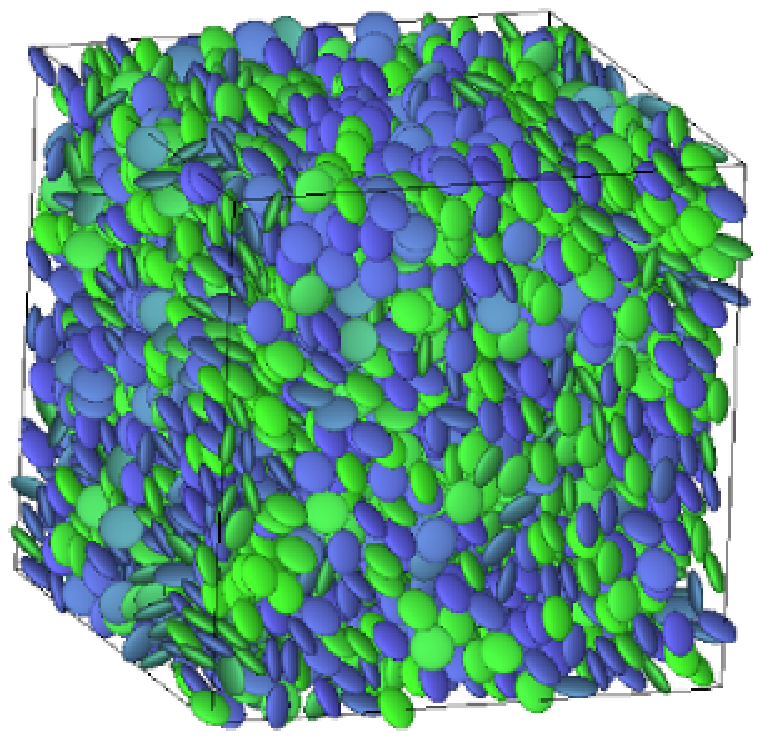}}\hspace{0.2cm}
\subfigure[\label{fig:p1c }]{\includegraphics[scale=0.6]{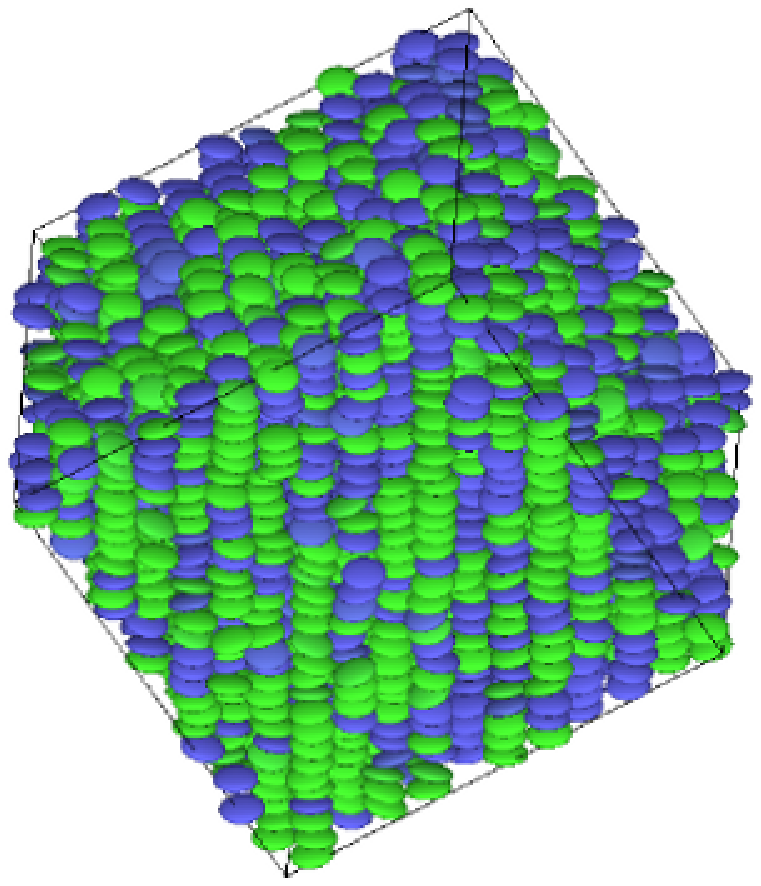}}\hspace{0.25cm}
\subfigure[\label{fig:p1d }]{\includegraphics[scale=0.4]{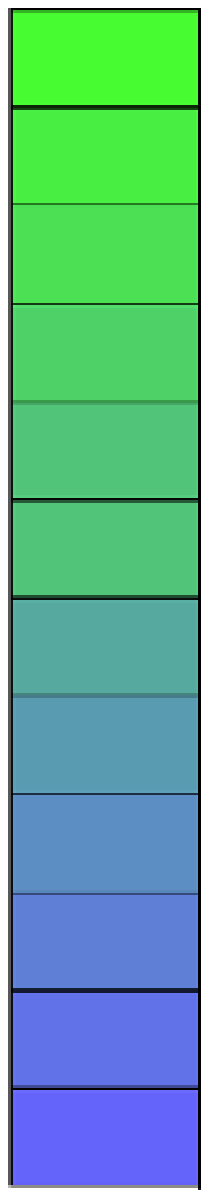}}

\caption{\label{fig:p1}(Color online). Snapshots of the configurations generated by MC
simulations at ( N=4000, \(\mu^{*}=0.20\)) : (a) Isotropic phase at \(T^{*}=7\), (b) Nematic phase at \(T^{*}=6\) , (c) Side view of the Hexagonal Columnar phase at \(T^{*}=5\), (d) The color plate showing different colors used in the snapshots given in this paper. The oblate ellipsoids are color coded according to their orientation with respect to the phase director ranging from parallel [ green (light gray)] to antiparallel [ blue (dark gray)] and the intermediate colors indicate intermediate orientations. All the snapshots are generated using the graphics software QMGA \cite{b38}.
 }

\end{figure*}
\begin{figure*}[b]
\centering
\subfigure[\label{fig:p8a}]{\includegraphics[scale=1.]{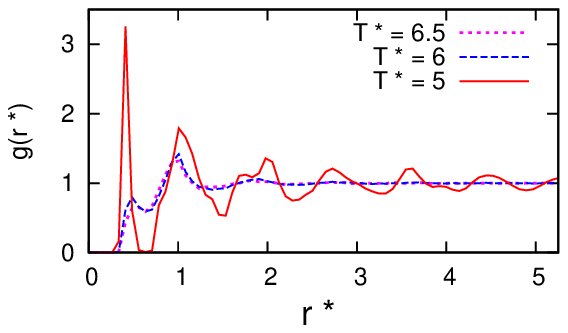}}
\subfigure[\label{fig:p8b}]{\includegraphics[scale=.8]{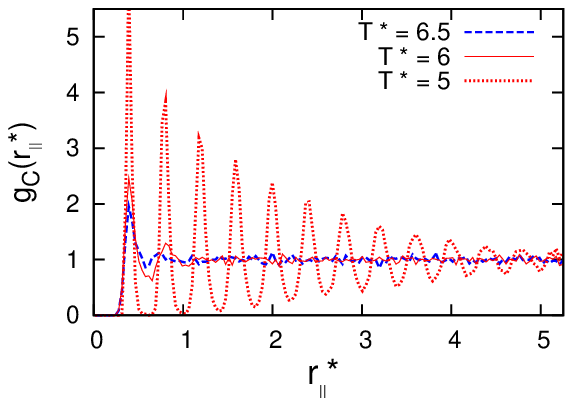}}
\subfigure[\label{fig:p8c}]{\includegraphics[scale=.8]{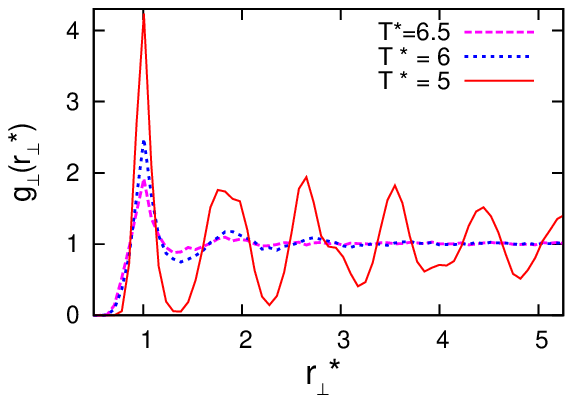}}
\caption{\label{fig:p8}(color online). Distribution functions for \(\mu^{*}\)=0.20 (N = 4000). (a) Radial distribution function \(g(r^{*})\) at three different temperatures: \(T^{*}=6.5\) (I) ,\(T^{*}=6.0\) (N) , \(T^{*}=5.0\) (Col); (b) Columnar distribution function \(g_{c}(r_{\parallel}^{*})\) at three different temperatures: \(T^{*}=6.5\), \(T^{*}=6.0\) and \(T^{*}=5.0\); (c) Perpendicular distribution function \(g(r_{\perp}^{*})\) at three different temperatures: \(T^{*}=6.5\), \(T^{*}=6.0\) and \(T^{*}=5.0\). The symbols I stands for isotropic, N stands for nematic phase, Col stands for the columnar phase.
 }
\end{figure*}
\begin{figure*}[tp]
\centering
\subfigure[\label{fig:p2a}]{\includegraphics[scale=0.75]{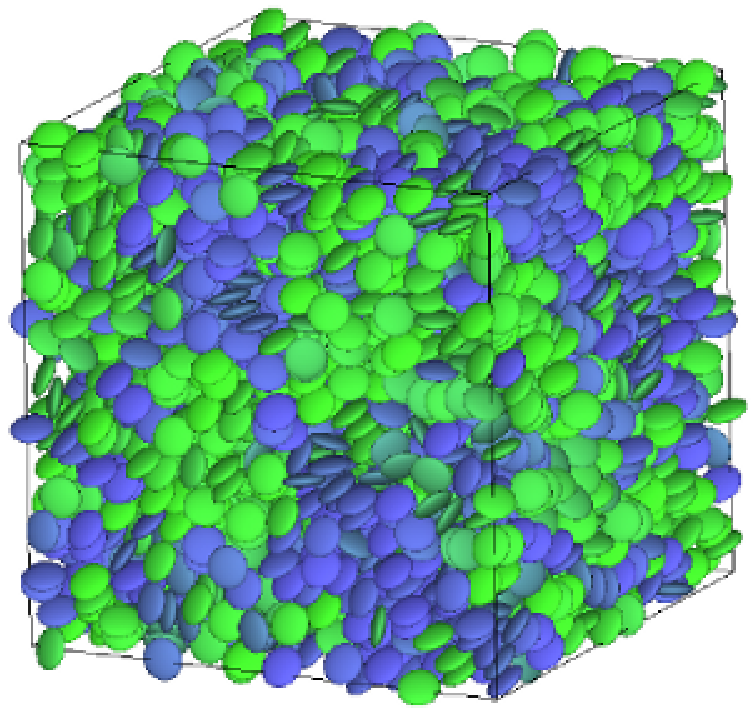}}\hspace{0.45cm}
\subfigure[\label{fig:p2b}]{\includegraphics[scale=0.7]{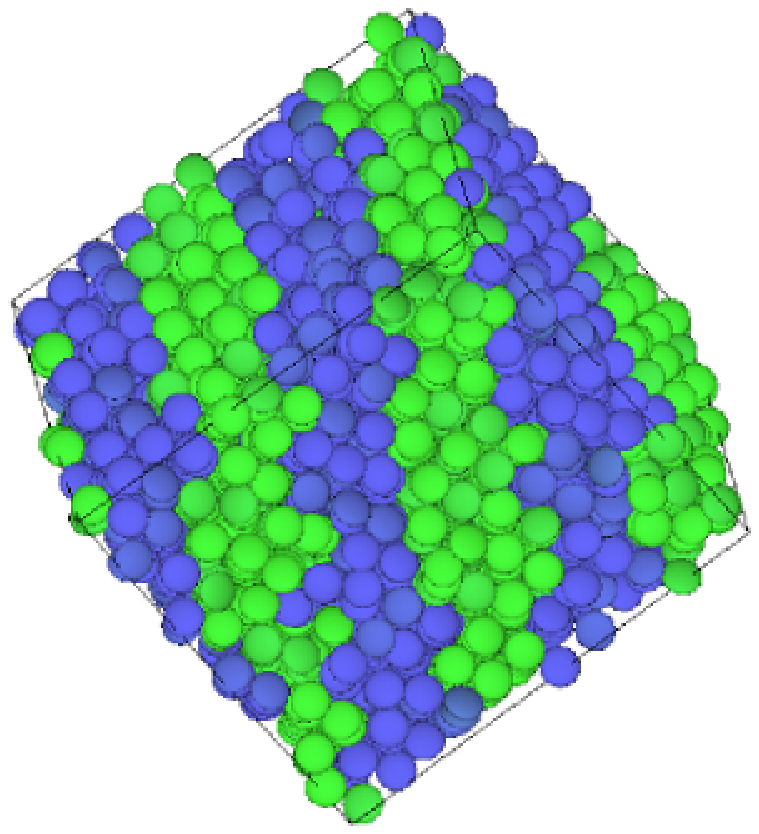}}\hspace{0.45cm}
\subfigure[\label{fig:p2c}]{\label{nematic}\includegraphics[scale=0.7]{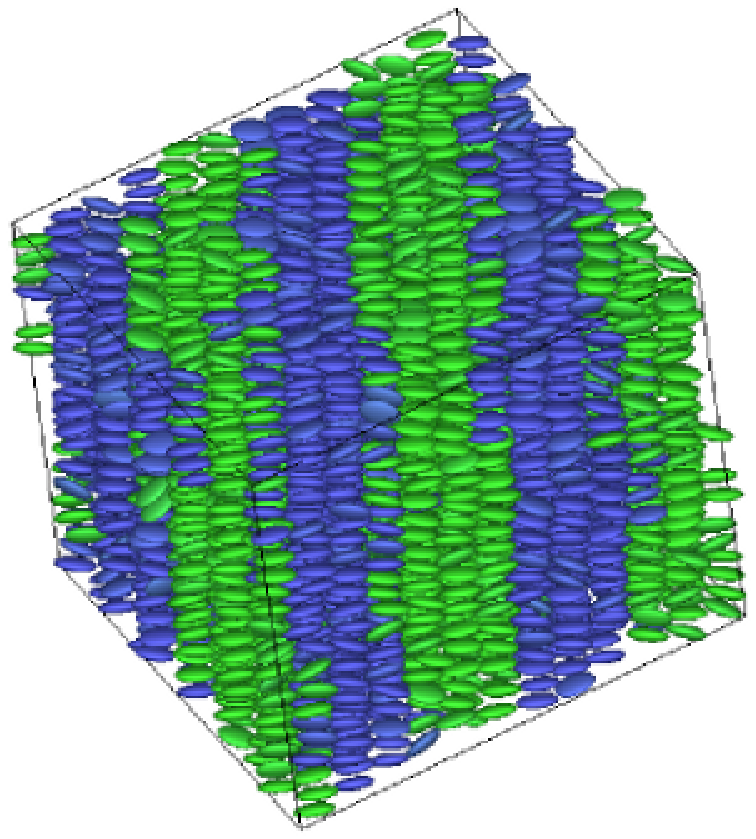}}\hspace{0.25cm}

\caption{\label{fig:p2}(color online). Snapshots of the configurations generated by MC
simulations at ( N=4000, \(\mu^{*}=0.40\)) : (a) Nematic phase at \(T^{*}=6.5\) , (b) Top view of the striped Hexagonal Columnar phase with periodically arranged domains of nearly equal and opposite polarizations at \(T^{*}=6\),(c) Side view of the striped Hexagonal Columnar phase at \(T^{*}=6\). The particles are color coded according to their orientation with respect to the phase director ranging from parallel [ green (light gray)] to antiparallel [ blue (dark gray)]. In the columnar phase, most ellipsoids are colored either in green (light gray) or blue(dark gray) to indicate that most of them are oriented either parallel or antiparallel to the director. The different polarized domains can be easily distinguished by their colors. 
 }

\end{figure*}

\begin{figure*}
\centering

\subfigure[\label{fig:p9a}]{\includegraphics[scale=1.]{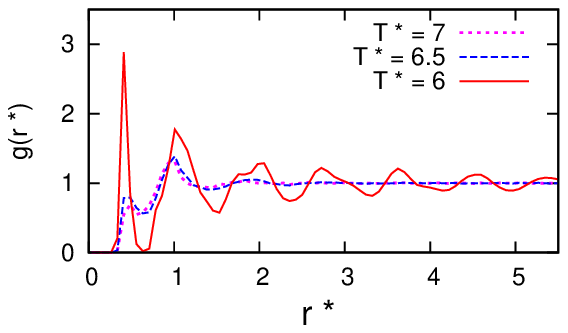}}
\subfigure[\label{fig:p9b}]{\includegraphics[scale=.8]{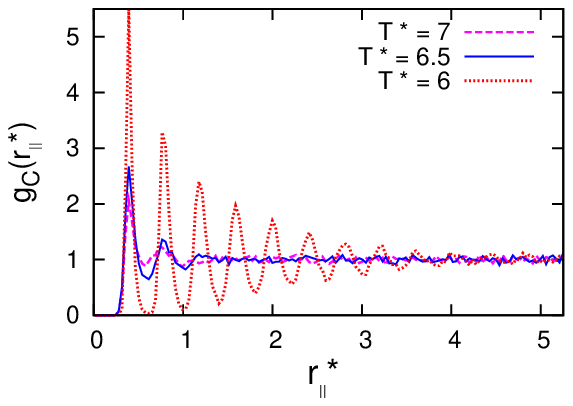}}
\subfigure[\label{fig:p9c}]{\includegraphics[scale=.8]{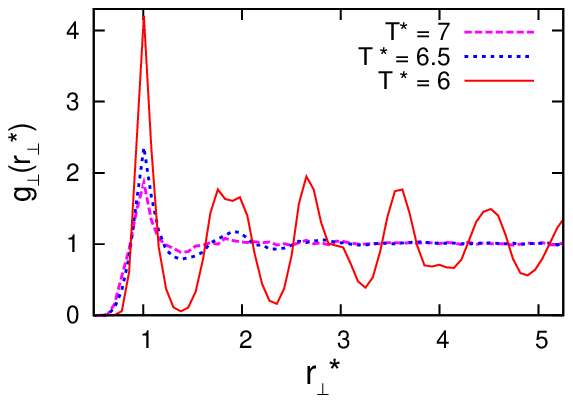}}

\caption{\label{fig:p9}(color online). Distribution functions for \(\mu^{*}\)=0.40 (N = 4000).(a) Radial distribution function \(g(r^{*})\) at three different temperatures: \(T^{*}=7\) (I) , \(T^{*}=6.5\) (N),\(T^{*}=6.0\) (SCol); (b) Columnar distribution function \(g_{c}(r_{\parallel}^{*})\) at three different temperatures: \(T^{*}=7\), \(T^{*}=6.5\) and \(T^{*}=6\); (c) Perpendicular distribution function \(g(r_{\perp}^{*})\) at three different temperatures: \(T^{*}=7\), \(T^{*}=6.5\) and \(T^{*}=6\). The symbols I stands for isotropic, N stands for nematic phase, SCol stands for the striped columnar phase. 
 }
\end{figure*}

\begin{figure*}
\centering

\subfigure[\label{fig:p10a}]{\includegraphics[scale=1.]{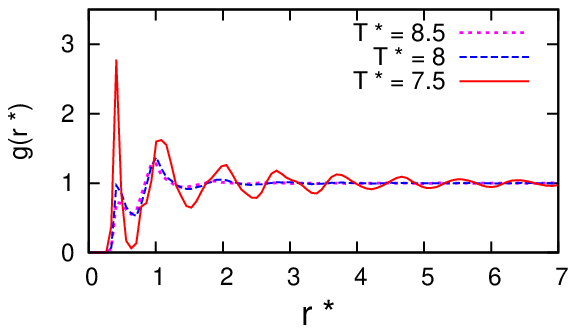}}
\subfigure[\label{fig:p10b}]{\includegraphics[scale=.8]{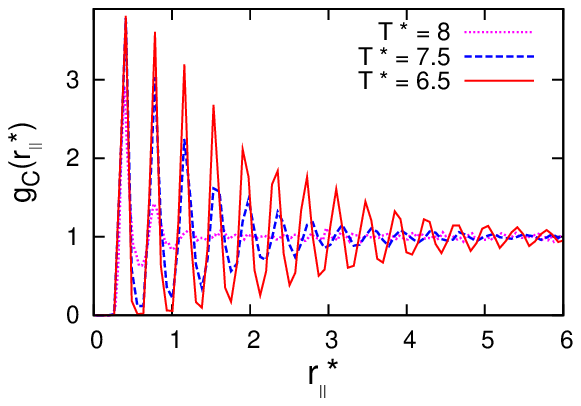}}
\subfigure[\label{fig:p10c}]{\includegraphics[scale=.8]{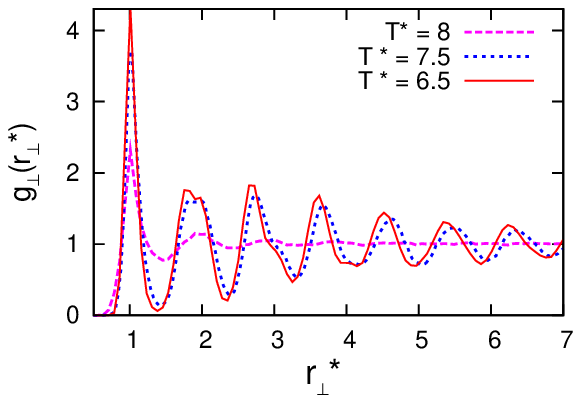}}

\caption{\label{fig:p10}(color online). Distribution functions for \(\mu^{*}\)=0.60 (N = 8500).(a) Radial distribution function \(g(r^{*})\) at three different temperatures: \(T^{*}=8.5\) (I) , \(T^{*}=8\) (SN),\(T^{*}=7.5\) (SCol); (b) Columnar distribution function \(g_{c}(r_{\parallel}^{*})\) at three different temperatures: \(T^{*}=8.0\), \(T^{*}=7.5\),\(T^{*}=6.5\) (SCol); (c) Perpendicular distribution function \(g(r_{\perp}^{*})\) at four different temperatures: \(T^{*}=8.0\), \(T^{*}=7.5\) and \(T^{*}=6.5\). The symbols I stands for isotropic, SN stands for the striped nematic phase, SCol stands for the striped columnar phase.
 }
\end{figure*}
\begin{figure*}
\centering
\hspace*{-0.525cm}\subfigure[\label{fig:p3a}]{\includegraphics[scale=0.6]{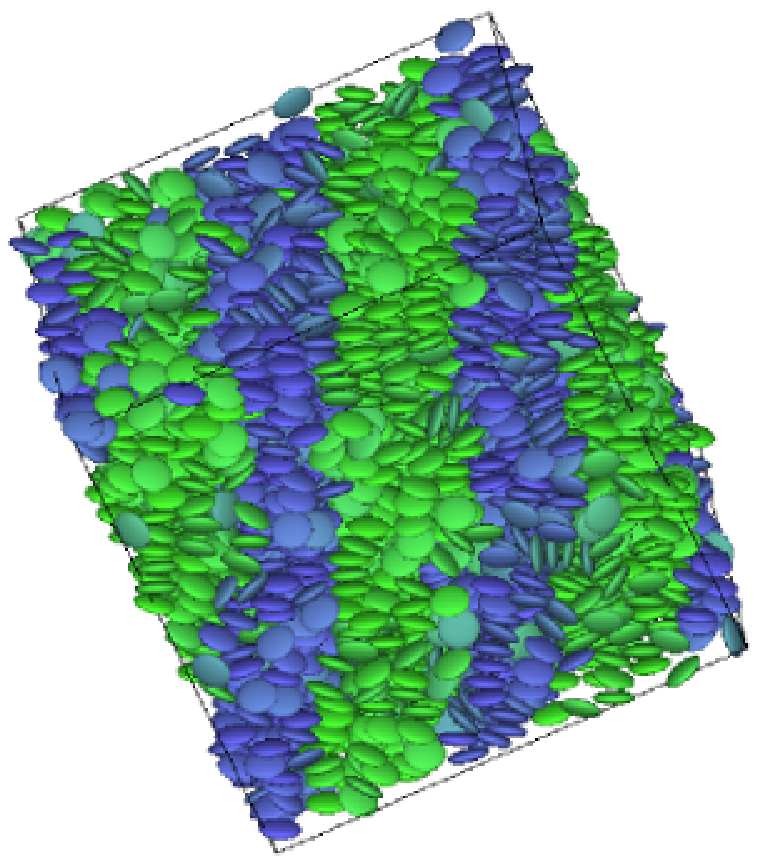}}
\subfigure[\label{fig:p3b}]{\label{nematic}\includegraphics[scale=0.7]{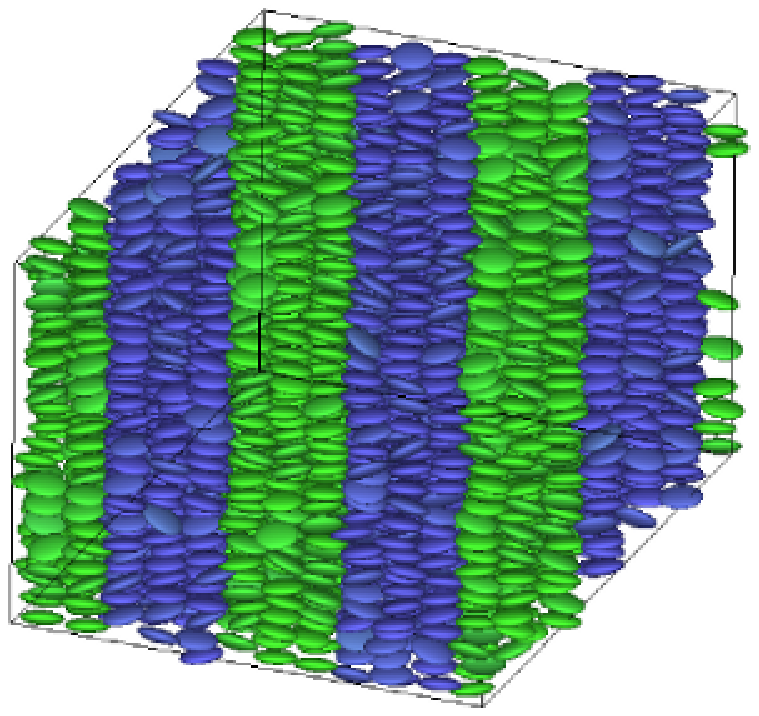}}
\subfigure[\label{fig:p3c}]{\includegraphics[scale=0.6]{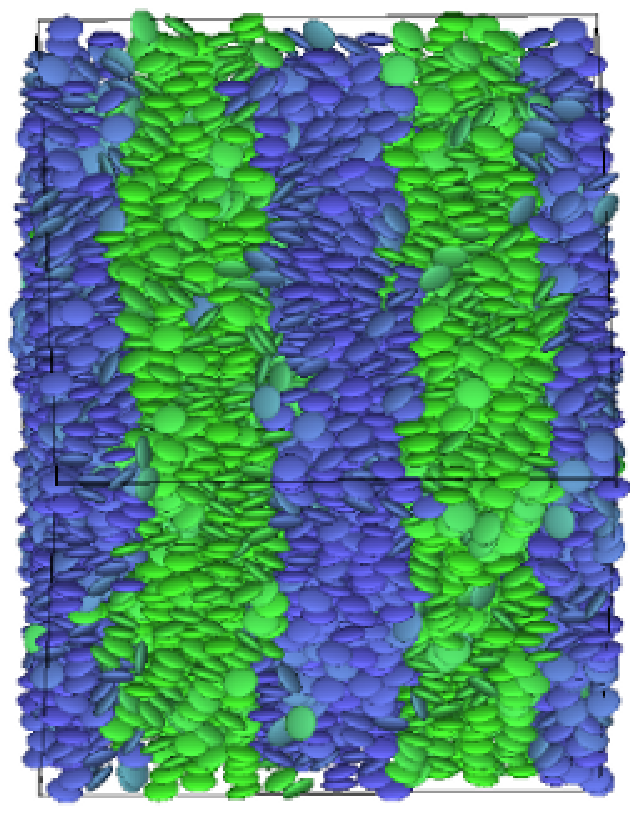}}\hspace{0.25cm}
\subfigure[\label{fig:p3d}]{\includegraphics[scale=0.6]{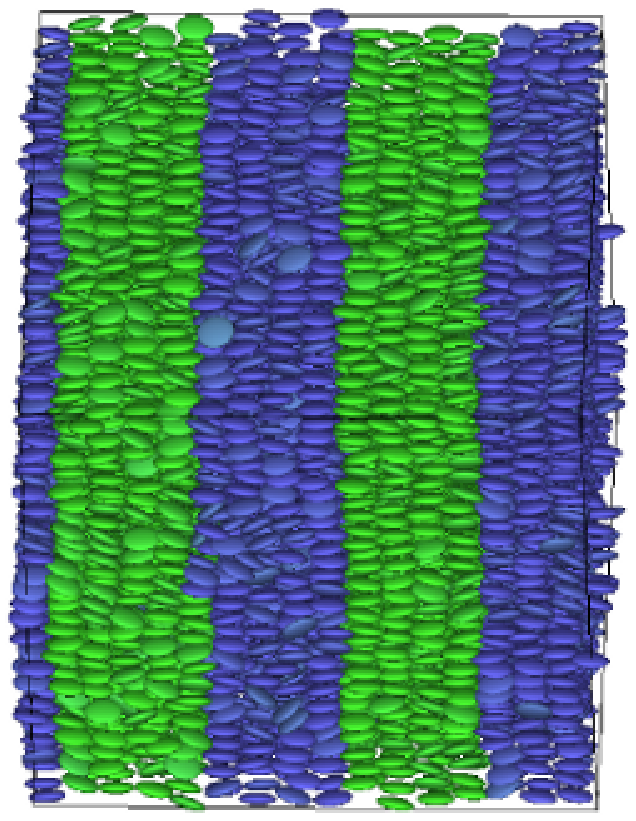}}\hspace{0.25cm}

\caption{\label{fig:p3}(color online). Snapshots of the configurations generated by MC
simulations for \(\mu^{*}=0.60\) : (a) striped nematic phase with polarized domains at \(T^{*}=8\) for N = 4000, (b) side view of the columnar phase with polarized domains at \(T^{*}=7.5\) for N = 4000, (c) striped nematic phase with polarized domains at \(T^{*}=8\) for N = 8500, (d) Side view of the striped Hexagonal Columnar phase at \(T^{*}=7\) for N=8500. The oblate ellipsoids are color coded according to their orientation with respect to the phase director ranging from parallel [ green (light gray)] to antiparallel [blue (dark gray)].  
 }

\end{figure*}
\begin{figure*}
\centering

\subfigure[\label{fig:p4a}]{\includegraphics[scale=0.5]{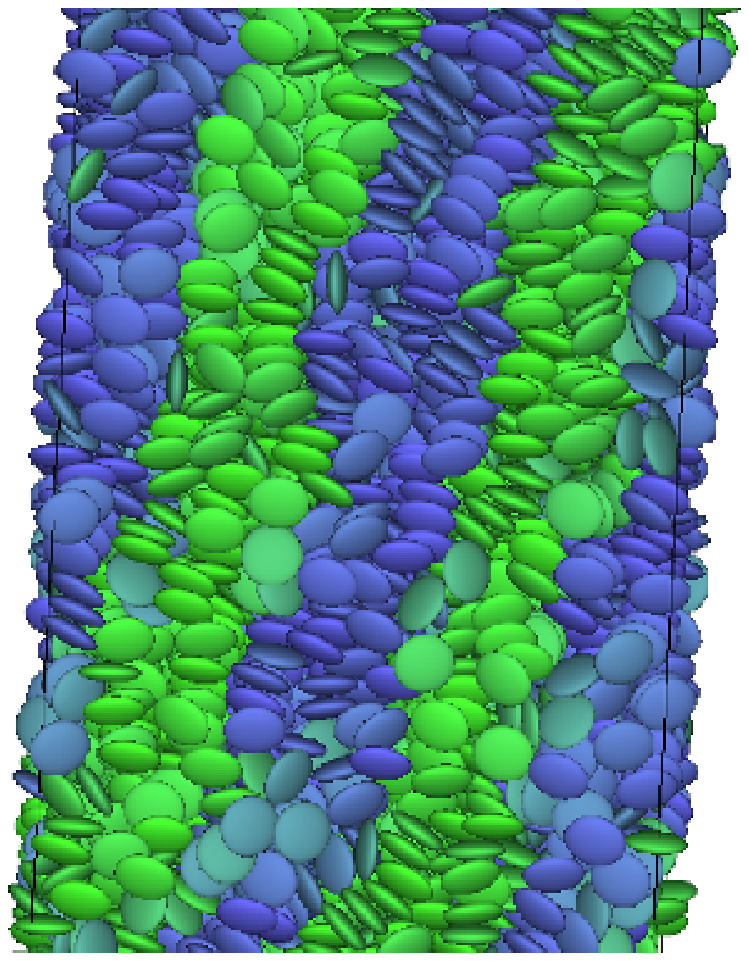}}\hspace{0.25cm}
\subfigure[\label{fig:p4b}]{\label{nematic}\includegraphics[scale=0.55]{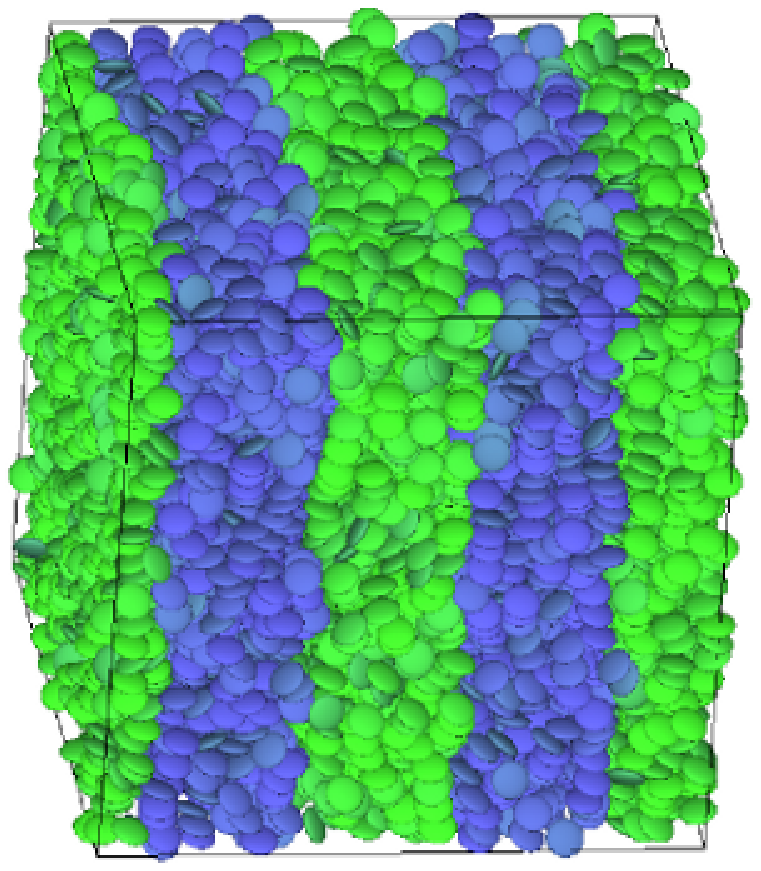}}\hspace{0.25cm}
\subfigure[\label{fig:p4c}]{\includegraphics[scale=0.6]{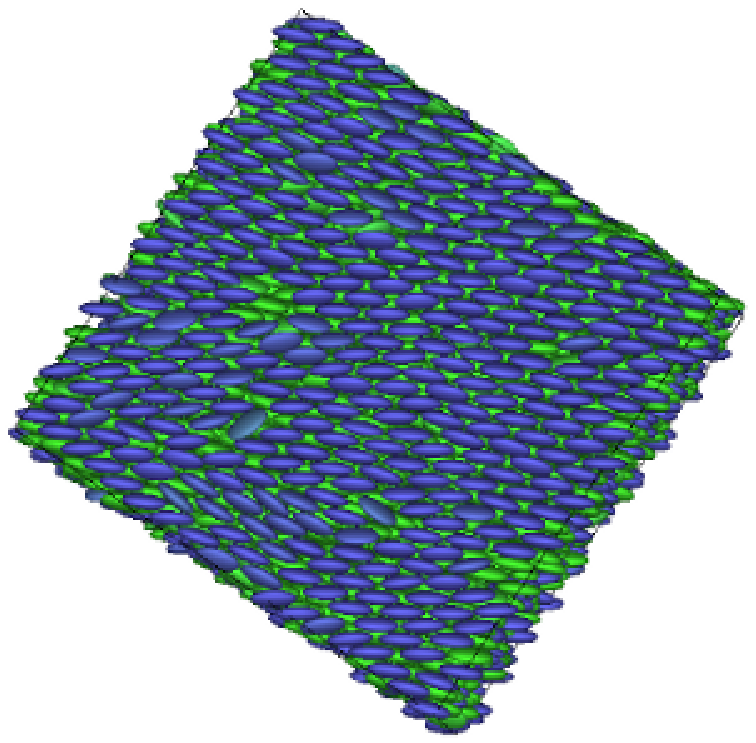}}\hspace{0.25cm}
\subfigure[\label{fig:p4d}]{\includegraphics[scale=0.5]{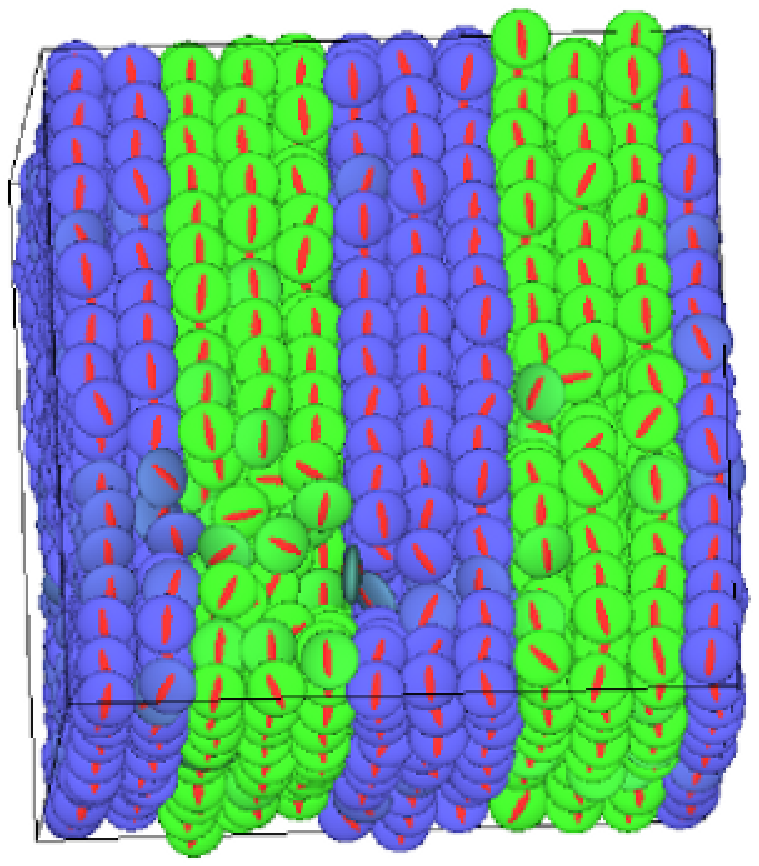}}

\caption{\label{fig:p4}(color online).Snapshots of the interesting configurations generated by MC
simulations for \(\mu^{*}=0.90\) : (a)  striped nematic phase with polarized domains at \(N=4000, T^{*}=11.5\),(b) striped nematic phase with polarized domains at \(N=8500, T^{*}=12\), (c) Side view of the striped biaxial phase with domains of equal and antiparallel polarization at \(N=4000, T^{*}=11\),(d) Top view of the striped biaxial phase at \(N=4000, T^{*}=11\) where the orientations of the dipolar separation vectors of each ellipsoid is indicated by red lines. The oblate ellipsoids are color coded according to their orientation with respect to the phase director ranging from parallel [ green (light gray)] to antiparallel [ blue (dark gray)].
 }

\end{figure*}

\begin{figure*}[!]
\centering
\subfigure[\label{fig:p13a}]{\includegraphics[scale=1.]{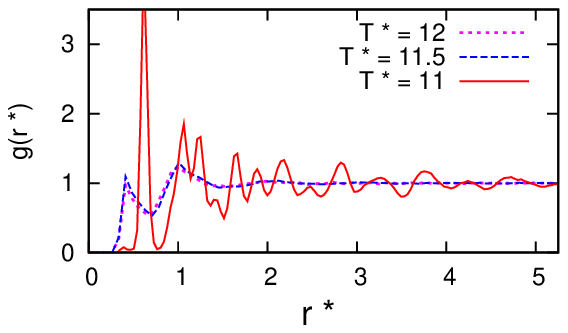}}
\subfigure[\label{fig:p13b}]{\includegraphics[scale=.8]{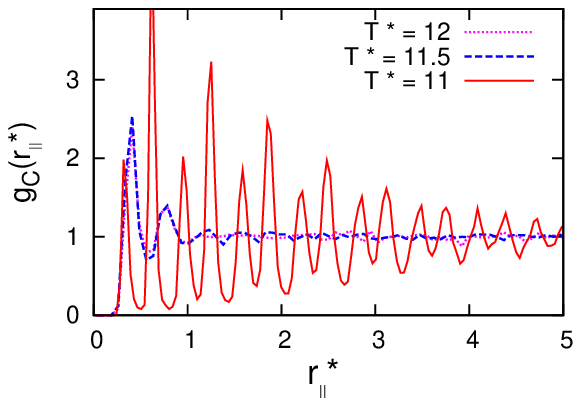}}
\subfigure[\label{fig:p13c}]{\includegraphics[scale=.8]{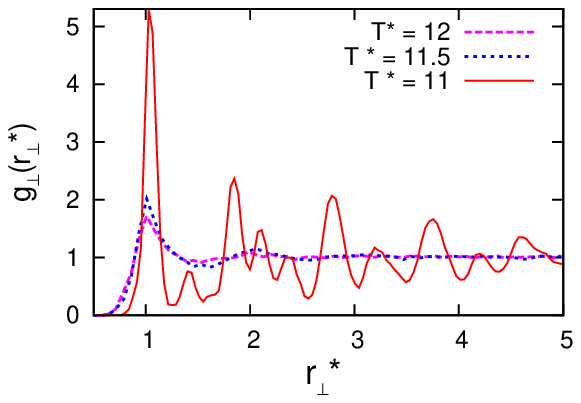}}

\caption{\label{fig:p13}(color online).Distribution functions for \(\mu^{*}\)=0.90 (N = 4000). (a) Radial distribution function \(g(r^{*})\) at three different temperatures: \(T^{*}=12\) (I) , \(T^{*}=11.5\) (SN),\(T^{*}=11.0\) (SB); (b) Columnar distribution function \(g_{c}(r_{\parallel}^{*})\) at three different temperatures: \(T^{*}=12\), \(T^{*}=11.5\) and \(T^{*}=11.0\); (c) Perpendicular distribution function \(g(r_{\perp}^{*})\) at three different temperatures: \(T^{*}=12\), \(T^{*}=11.5\) and \(T^{*}=11.0\). The symbols SN stands for the striped nematic and SB stands for the striped biaxial phase.
 }
\end{figure*}
\subsection{Results obtained using the spherical cutoff approach}

For \( \mu^{*}=0.20 \), the strength of dipolar interaction remains quite weak to significantly influence the phase behavior of the GB ellipsoids. The system in this case exhibits a phase behavior qualitatively similar to that of the non-polar GB disks as reported in \cite{b282}. The variations of the average orientational order parameter \(\langle P_{2}\rangle\) against the reduced temperature are presented in  Fig. \ref{fig:p7a} for two systems of different size. It can be seen that the behavior remain quite similar for N=1500 and N=4000. \(\langle P_{2}\rangle\) remains close to zero at high temperatures identifying the isotropic phase. Upon cooling \(\langle P_{2}\rangle\) jumps to \(\langle P_{2}\rangle \approx0.71\) at \(T^{*}=6\) indicating a transition to the orientationally ordered discotic nematic phase. At \(T^{*}=5\), we observe a discotic columnar phase without any well defined polar arrangement of the ellipsoids. The snapshots of these resultant configurations are shown in Fig. \ref{fig:p1} and the related distribution functions are presented in Fig. \ref{fig:p8} for N=4000. The flatness in the radial distribution function \(g(r^{*})\) at \(T^{*}=6.5\) reflects the structurelessness of the isotropic liquid in the long range. At \(T^{*}=6\), \(g(r^{*})\) shows the absence of any long-range positional order in the nematic phase. Considerable structure in \(g(r^{*})\) for \(T^{*}=5\) indicates the formation of the higher ordered columnar structures. The columnar distribution function \(g_{c}(r_{\shortparallel}^{*})\), which is a measure of positional order within a single column, is shown in  Fig. \ref{fig:p8b}. The periodic nature of \(g_{c}(r_{\shortparallel}^{*})\) confirms the periodic stacking of molecules in the columnar phases. At \(T^{*}=5\), \(g_{c}(r_{\shortparallel}^{*})\) decays algebraically, indicating a quasi longrange columnar order. The algebraic decay of the peak values in the columnar
distribution function is a typical signature of columnar liquid crystal order \cite{b30}. For a crystal, the peak intensities remain finite at long range, which is clearly not the case. As the temperature is lowered, \(g_{c}(r_{\parallel}^{*})\) exhibits more ordered structure. The small peaks for \( r^{*}, r_{\shortparallel}^{*} <0.5\) in \(g_{c}(r_{\shortparallel}^{*})\) and in \(g(r^{*})\) describes finite probability of short range face-to-face ordering in the nematic phase.  Fig. \ref{fig:p8c} shows the perpendicular distribution function \(g(r_{\perp}^{*})\), which is a measure of translational order in the plane orthogonal to the orientation of the disc shaped molecules. \(g(r_{\perp}^{*})\) at \(T^{*}=6.0\) remains essentially flat for the nematic phase. At \(T^{*}=5\), a flat second peak indicates hexagonal columnar packing. The second peak is usually broken for for a perfect hexagonal order. The biaxial order parameter \( \langle R_{2,2}^{2}\rangle \) and the global ferroelectric order parameter \(\langle P_{1}\rangle \) remain \(\approx 0\) in all the above described phases for \(\mu^{*}=0.20\).\\

For \(\mu^{*}=0.40\), the variations of the average orientational order parameter \(\langle P_{2}\rangle\) against the reduced temperatures are shown in Fig. \ref{fig:p7b} for N=1500 and N=4000. The isotropic liquid condenses to a discotic nematic phase without any well defined polar ordering at \(T^{*}=6.5\) with \(\langle P_{2}\rangle\ \approx0.66\). 
  For \(T^{*} \leq 6.0\), the systems generate interesting columnar structures consisting of periodic array of ferroelectric slablike domains arranged in an antiferroelectric fashion. The striped columnar phase at \(T^{*}=6\) is shown in  Figs. \ref{fig:p2b} and \ref{fig:p2c}. It can be clearly seen that the system has split into a number of ferroelectric domains with alternating polarization in the columnar phase. The colors indicate two mutually anti parallel directions of polarization of the domains. Each domain consists of a number of axially polarized columns of the ellipsoids. The related columnar and transverse distribution functions are shown in  Figs. \ref{fig:p9b} and \ref{fig:p9c}. The formation of anti parallel polarized domains can be qualitatively understood as follows. If a spherical ferroelectric sample is surrounded by vacuum, then charges induced on the surface of the sphere creates a depolarizing electric field with energy proportional to the macroscopic polarization. To minimize the configuration energy, domains are generated with opposite polarization which destroy the overall polarization. There is a competition between the energy decrease due to domain formation and energy increase due to domain wall formation. The domains are formed until these two balances each other. A similar scenario can be observed in case of ferroelectric crystals \cite{b36} and in model systems of dipolar spheres \cite{b21,b22}. Here, the striped phase shows a number of distinct ferroelectric domains with alternating polarization rather than different domains with different directions of polarization found for dipolar spheres \cite{b22}. In addition the domains are characterized with a strong axial polarization. At \(T^{*}=6.5\), the radial distribution function \(g(r^{*})\) in  Fig. \ref{fig:p9a} shows the absence of any long-range positional order in the nematic phase. Considerable structure in \(g(r^{*})\) for \(T^{*}=6\) indicates the formation of higher ordered striped columnar structure. At \(T^{*}=6\), \(g_{c}(r_{\shortparallel}^{*})\) decays rapidly, indicating a quasi long-range order along column axis. As the temperature is lowered, \(g_{c}(r_{\parallel}^{*})\) exhibited more ordered structure.  Fig. \ref{fig:p9c} shows the perpendicular distribution function \(g(r_{\perp}^{*})\), which is a measure of translational order in the plane orthogonal to the orientation of the disc shaped molecules. \(g(r_{\perp}^{*})\) at \(T^{*}=6.5\) remains essentially flat for the nematic phase. At \(T^{*}=6.0\), the broken second peak indicates proper hexagonal columnar packing. 
 The biaxial order parameter \( \langle R_{2,2}^{2}\rangle \) and the global ferroelectric order parameter \(\langle P_{1} \rangle \) remain \(\approx 0\) in all the above described phases for \(\mu^{*}=0.40\).\\
\begin{figure}[!]
\centering
\includegraphics[scale=1.0]{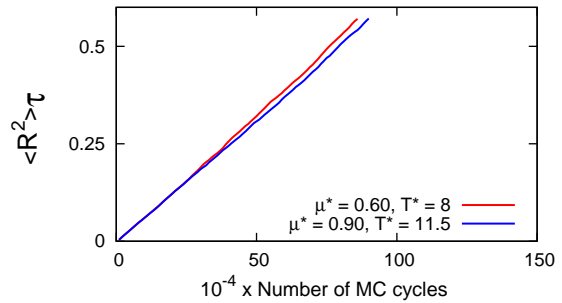}

\caption{\label{fig:p33}(color online). Mean-squared displacement \(\langle R^{2} \rangle_{\tau}\) against MC cycles for the striped nematic phases at (\(T^{*}=8,\mu^{*}=0.60\)) and (\(T^{*}=11.5,\mu^{*}=0.90\)) for the systems of \textit{N} = 4000 dipolar particles. }

\end{figure}

For \(\mu^{*}=0.60\), the system directly transforms from an isotropic liquid to a striped nematic fluid at \(T^{*}=8\) for N = 4000 and N = 8500.  However a smaller system (N = 1500) does not exhibit the striped nematic phase at \(T^{*}=8\). The snapshots of the striped nematic configurations are shown in Figs. \ref{fig:p3a} and \ref{fig:p3c} respectively for N = 4000 and N = 8500. The fascinating periodic modulation of ferroelectric order obtained in the striped nematic phase is similar to that observed in the striped columnar phases found for \(\mu^{*}=0.40\). The snapshots in Fig. \ref{fig:p3} clearly show that systems have generated slablike ferroelectric domains periodically arranged in an antiferroelectric fashion. Different colors have been used to indicate different domains of approximately antiparallel overall polarization. The striped columnar phase appears at \(T^{*}=7.5\) for all the system sizes. The similarity in the phase behavior of the two larger systems can be seen from Fig. \ref{fig:p7c} describing the variation of the average orientational order parameter with the reduced temperature. The snapshots also indicate that each domain is strongly polarized. \\ 

The snapshots in Fig. \ref{fig:p3} show that the domain size increases with the system size N in case of the striped nematic and columnar phases. For N=1500, each polarized slablike domain in the striped columnar phases consists of two rows and for N = 4000 and N=8500, the domains clearly consist of three and four rows respectively as can be seen from figs.\ref{fig:p3b} and \ref{fig:p3d}. The gradual increase in the domain size clearly indicates that the consideration of longer range of dipolar interaction favors the ferroelectric order. In addition, the snapshots in Fig.\ref{fig:p3} clearly demonstrate the absence of any long range structure in the striped nematic phases. In order to verify the fluidity of the nematic phase, we calculated the mean square displacement \( \langle R^{2} \rangle_{\tau}\) and the corresponding plot is shown in  Fig. \ref{fig:p33}. The mean square displacement steadily increased with increasing \(\tau\) indicating highly fluid behavior at \(T^{*}=8\). The related structural distribution functions for \(\mu^{*}=0.60, N=8500\) are shown in  Fig. \ref{fig:p10}.  At \(T^{*}=8\), the radial distribution function \(g(r^{*})\) in  Fig. \ref{fig:p10a} shows the absence of any long-range positional order in the nematic phase. Considerable structure in \(g(r^{*})\) for \(T^{*}\le7.5\) indicates the formation of higher ordered columnar structures. At \(T^{*}=7.5\), \(g_{c}(r_{\shortparallel}^{*})\) decays rapidly, indicating a quasi long-range order along column axis. The snapshot in Figs.\ref{fig:p3b}and \ref{fig:p3d} show the positional order along the columns of the columnar fluid structures. Fig. \ref{fig:p10c} shows the perpendicular distribution function \(g(r_{\perp}^{*})\) which at \(T^{*}=8\) remains essentially flat for the nematic phase. At \(T^{*}=6.5\), the broken second peak indicates proper hexagonal columnar packing. The biaxial order parameter \( \langle R_{2,2}^{2}\rangle \) and the global ferroelectric order parameter \( \langle P_{1} \rangle \) remain \(\approx 0\) in all the above described phases for \(\mu^{*}=0.60\)\\

Let us now describe the phase behavior for \(\mu^{*}\) = 0.90. The system exhibits a striped biaxial phase having slablike ferroelectric domains in addition to the uniaxial striped nematic phase. The phase biaxiality originates exclusively from the strong dipolar interactions. No columnar phase is obtained at this dipole strength. The variations of the orientational order parameter \(\langle P_{2} \rangle\) against the reduced temperatures for different system sizes are shown in Fig. \ref{fig:p7d}. It can be seen from Figs. \ref{fig:p7c} and \ref{fig:p7d} that the temperature range of the striped nematic phase gradually increases with the dipole strength for a fixed system size which is similar to the behavior of the system when studied under the influence of the conducting boundary conditions in our previous investigation \cite{b4444} where the temperature range of the single domain ferroelectric nematic phase gradually increased with \(\mu^{*}\). Here, a system size effect is also observed as the temperature range of the striped nematic phase is found to increase with N as indicated by the plots in \ref{fig:p7d}. The largest system consisting of N = 8500 dipolar ellipsoids exhibits a striped nematic phase at \(T^{*}=12\) with \( \langle P_{2} \rangle \approx 0.5\). The ordering of ferroelectric domains as shown in Figs. \ref{fig:p4a} and \ref{fig:p4b} is quite similar to that observed for the striped nematic and striped columnar phases obtained for lower dipole strengths. 
 Fig. \ref{fig:p13a} shows the radial distribution function \(g(r^{*})\) at different temperatures. The flatness in \(g(r^{*})\) at \(T^{*}=11.5\) reflects the structurelessness of the striped nematic liquid in the long range. At \(T^{*}=11.5\), \(g_{c}(r_{\shortparallel}^{*})\) and \(g_{\perp}(r_{\perp}^{*})\) also show no sign of any long range positional order.  We have also calculated the mean square displacement \( \langle R^{2} \rangle_{\tau}\) for the striped nematic phase at \(T^{*}=11.5\). The mean square displacement steadily increased with increasing \(\tau\) indicating fluid behavior as shown in Fig. \ref{fig:p33}. Considerable structure in \(g(r^{*})\) at \(T^{*}=11\) indicates the formation of a more ordered biaxial phase. We have measured the conventional biaxial order parameter \( \langle R_{2,2}^{2}\rangle \), the value which is \(\approx\) 0.95 as measured in the biaxial phases. However, \( \langle R_{2,2}^{2}\rangle \) remain \(\approx\) 0 in the striped nematic phases reported here. The columnar and transverse distribution functions shown in  Figs. \ref{fig:p13b} and \ref{fig:p13c}, are quite different from those obtained for the columnar phases. From the snapshots of the biaxial phases, we can understand the related structure. From  Fig. \ref{fig:p4d}, it can be seen that the dipoles are arranged such that the dipolar separation vectors pointing from one dipole on a disc to the other get oriented mostly in same direction. This ordering can be understood as the effect of strong pair interaction possible in this condition between two terminal dipoles of neighboring ellipsoids if two parallel dipoles are placed above each other. The snapshot of such intercalated arrangement is shown in Fig. \ref{fig:p4c}. The structural distribution functions for the biaxial phase are shown in Fig. \ref{fig:p13a}-\ref{fig:p13c} At \(T^{*}=11\), \(g_{c}(r_{\shortparallel}^{*})\) indicates the intercalated arrangement of the ellipsoids. The shorter peaks generate due to the presence of the pairs of side-by-side ellipsoids covering single ellipsoids. It should be noted that strong dipoles stabilize striped nematic phases with relatively weak orientational order. The global ferroelectric order parameter\( \langle P_{1} \rangle\) remained approximately equal to zero indicating zero net polarization for all the phases observed for \(\mu^{*}=0.90\).
\subsection{Results obtained using the Ewald summation method}

We have also studied the phase behavior using the conventional but more time consuming Ewald Summation method with \(\epsilon_{s}=1\). 
We have chosen a smaller system consisting of N = 500 dipolar ellipsoids to study the depolarizing effects using ES approach and investigated the phase behaviors for two different dipole strengths \(\mu^{*}=0.70\) and \(\mu^{*}=0.90\) for which we observe the ferroelectric nematic, ferroelectric columnar and ferroelectric biaxial order within periodic slablike ferroelectric domains. 

\begin{figure*}[!]
\hspace*{-.72cm}
\subfigure[\label{fig:p21a}]{\includegraphics[scale=0.37]{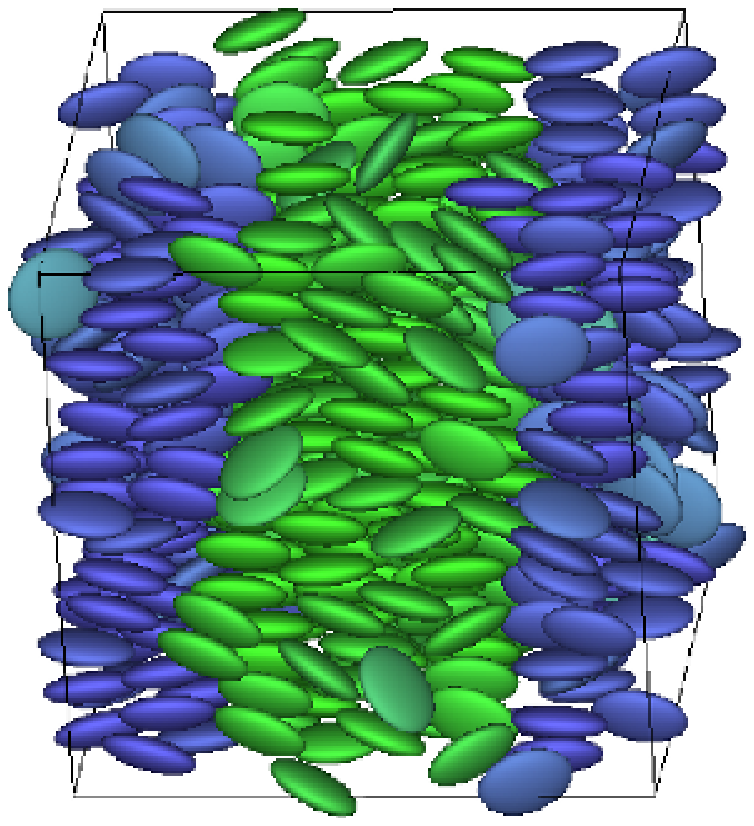}}
\subfigure[\label{fig:p21b}]{\label{nematic}\includegraphics[scale=0.4]{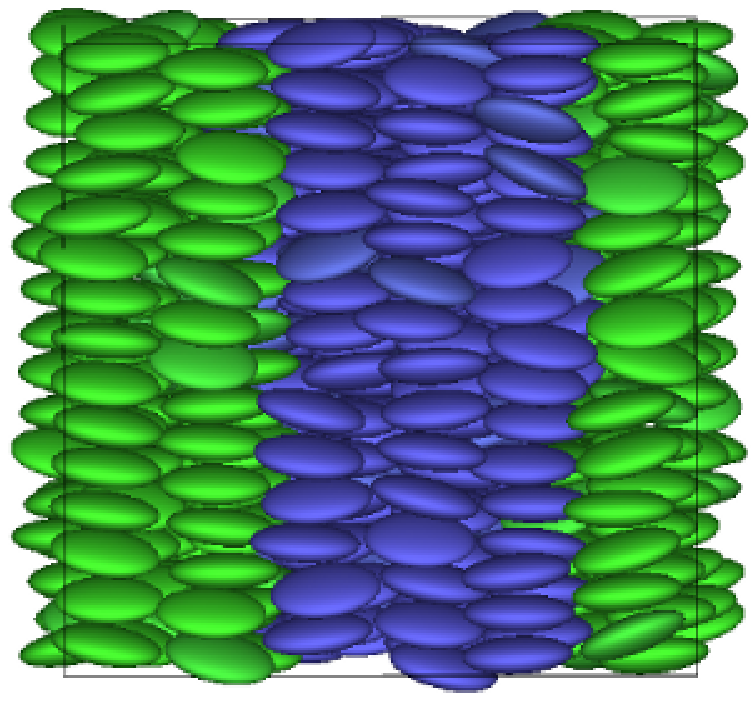}}
\subfigure[\label{fig:p21c}]{\includegraphics[scale=0.37]{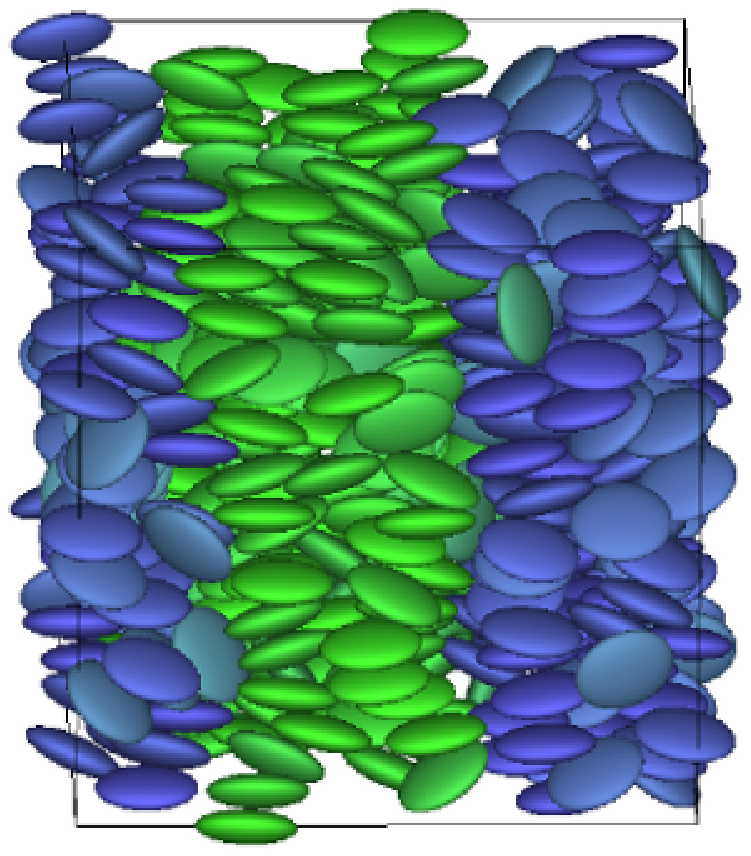}}
\subfigure[\label{fig:p21d}]{\includegraphics[scale=0.43]{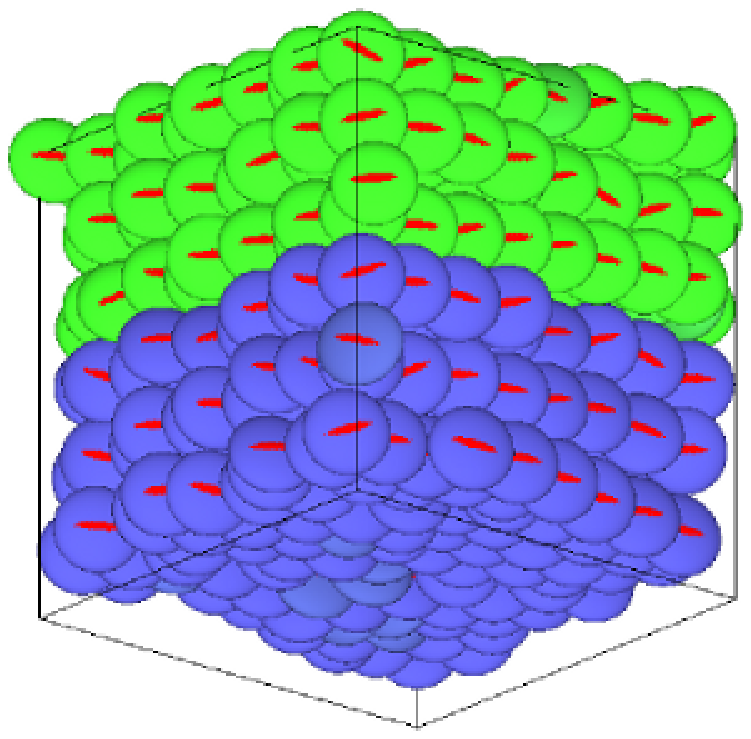}}
\subfigure[\label{fig:p21a1}]{\includegraphics[scale=0.8]{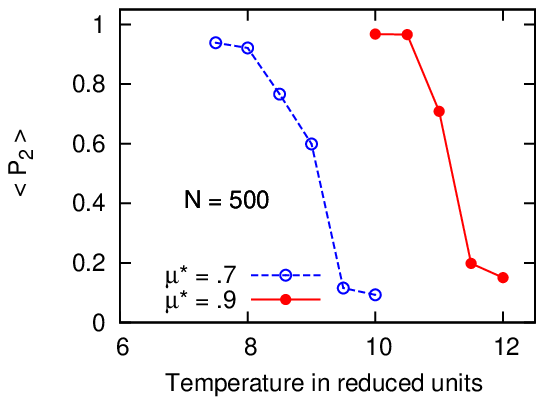}}

\caption{\label{fig:p212}(Color online). Snapshots of the configurations generated by MC
simulations at N = 500 using Ewald Summation method : (a) striped nematic phase at \(\mu^{*}=0.7,T^{*}=8.5\), (b) Striped columnar phase at \(\mu^{*}=0.7,T^{*}=7.5\) , (c) Striped nematic phase at \(\mu^{*}=0.9,T^{*}=11\), (d)  Striped biaxial phase at \(\mu^{*}=0.9,T^{*}=10.5\). The oblate ellipsoids are color coded according to their orientation with respect to the phase director ranging from parallel [ green (light gray)] to antiparallel [ blue (dark gray)] and the intermediate colors indicate intermediate orientations. }

\end{figure*}

\begin{figure*}[!]
\centering

\subfigure[\label{fig:p88a}]{\includegraphics[scale=1.]{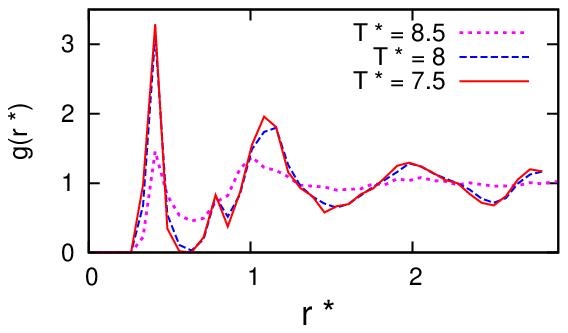}}
\subfigure[\label{fig:p88b}]{\includegraphics[scale=.8]{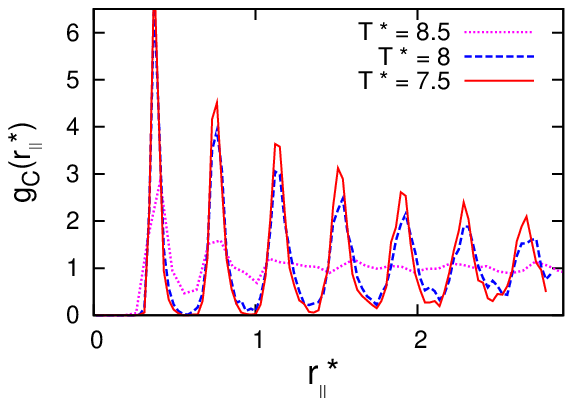}}
\subfigure[\label{fig:p88c}]{\includegraphics[scale=.8]{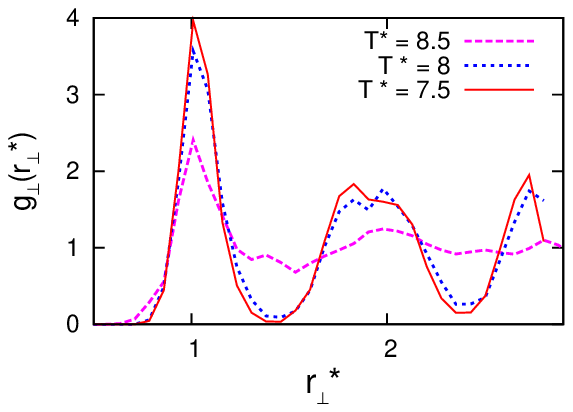}}

\caption{\label{fig:p88}(color online). Distribution functions for \(\mu^{*}\)=0.70 (N = 500). (a) Radial distribution function \(g(r^{*})\) at three different temperatures: \(T^{*}=8.5\) (SN) ,\(T^{*}=8.0\) (SN) and \(T^{*}=7.5\) (SCol); (b) Columnar distribution function \(g_{c}(r_{\parallel}^{*})\) at three different temperatures: \(T^{*}=8.5\),\(T^{*}=8.0\) and \(T^{*}=7.5\); (c) Perpendicular distribution function \(g(r_{\perp}^{*})\) at three different temperatures: \(T^{*}=8.5\), \(T^{*}=8.0\) and \(T^{*}=7.5\). The symbols SN stands for striped nematic phase, SCol stands for the striped columnar phase.
 }
\end{figure*}

For \(\mu^{*}=0.70\), the striped nematic phase with ferroelectric slablike domains is obtained at \(T^{*}=9\) upon cooling the isotropic phase at the fixed pressure. The snapshot in Fig. \ref{fig:p21a} clearly shows the presence of ferroelectric nematic domains with mutually opposite polarization at \(T^{*}=8.5\). Further lowering of temperature gives rise to a columnar liquid crystal, at \(T^{*}=8\), with ferroelectric domains consisting of three rows of columns. The plots of the structural distribution functions are presented in Fig. \ref{fig:p88}. The flatness in the radial distribution function \(g(r^{*})\) at \(T^{*}=8.5\) confirms the absence of any long-range positional order in the striped nematic phase shown in Fig. \ref{fig:p21a}. Considerable structure in \(g(r^{*})\) for \(T^{*}\le8\) indicates the formation of the higher ordered columnar structures. The columnar distribution function \(g_{c}(r_{\shortparallel}^{*})\) is shown in  Fig. \ref{fig:p88b}. At \(T^{*}=8\), \(g_{c}(r_{\shortparallel}^{*})\) decays algebraically, indicating a quasi longrange columnar order as shown in the snapshot in Fig.\ref{fig:p21b}. The small peaks for \( r^{*}, r_{\shortparallel}^{*} <0.5\) in \(g_{c}(r_{\shortparallel}^{*})\) and in \(g(r^{*})\) indicates finite probability of short range face-to-face ordering in the nematic configurations.  Fig. \ref{fig:p88c} shows the perpendicular distribution function at \(T^{*}=8.5\) remains essentially flat for the nematic phase. At \(T^{*}=8\), a broken second peak indicates hexagonal columnar packing of the columns. The above characteristics are qualitatively similar to that of the (N = 4000, \(\mu^{*}=0.60\)) system investigated using the SC technique. The biaxial order parameter \( \langle R_{2,2}^{2}\rangle \) and the global ferroelectric order parameter \(\langle P_{1} \rangle \) remain \(\approx 0\) in all the above described phases for \(\mu^{*}=0.70\).\\

\begin{figure*}[!]
\centering

\subfigure[\label{fig:p98a}]{\includegraphics[scale=1.]{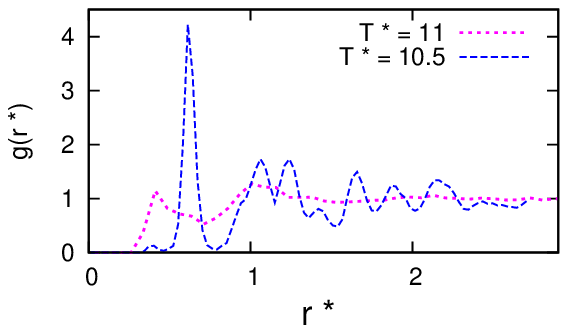}}
\subfigure[\label{fig:p98b}]{\includegraphics[scale=.8]{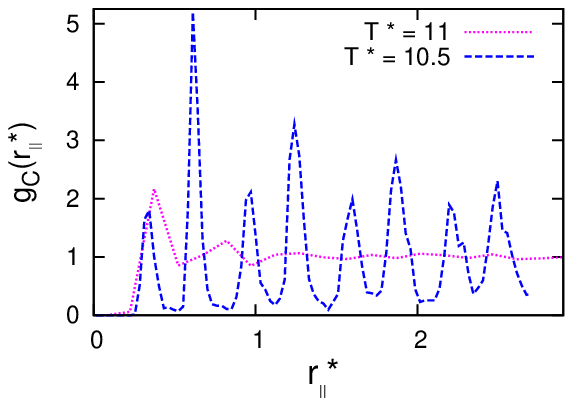}}
\subfigure[\label{fig:p98c}]{\includegraphics[scale=.8]{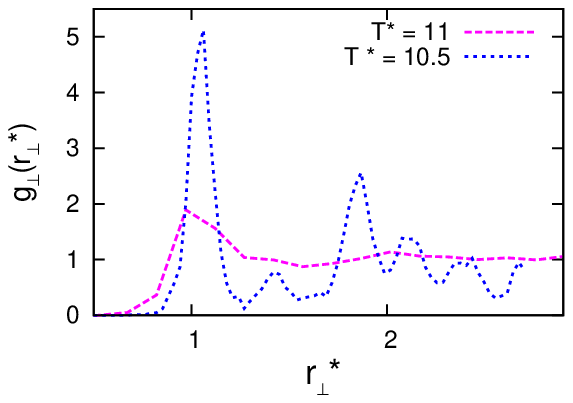}}

\caption{\label{fig:p98}(color online). Distribution functions for \(\mu^{*}\)=0.90 (N = 500). (a) Radial distribution function \(g(r^{*})\) at two different temperatures: \(T^{*}=11\) (SN) and \(T^{*}=10.5\) (SB); (b) Columnar distribution function \(g_{c}(r_{\parallel}^{*})\) at two different temperatures: \(T^{*}=11\) and \(T^{*}=10.5\); (c) Perpendicular distribution function \(g(r_{\perp}^{*})\) at two different temperatures: \(T^{*}=11\) and \(T^{*}=10.5\). The symbols SN stands for striped nematic phase, SB stands for the striped biaxial phase.
 }
\end{figure*}

For \(\mu^{*}=0.90\), the striped nematic phase having ferroelectric slablike domains is obtained at \(T^{*}=11\) upon cooling the isotropic phase. The variations of the average orientational order parameter \(\langle P_{2}\rangle\) against the reduced temperatures are presented in Fig. \ref{fig:p21a1} and the snapshots of the interesting configurations obtained using the Ewald sum method are given in Figs. \ref{fig:p21c}-\ref{fig:p21d}. The presence of ferroelectric nematic domains with antiparallel polarizations at the above temperature can be seen from the snapshots. Further lowering of temperature generates a highly ordered biaxial phase, at \(T^{*}=10.5\), with ferroelectric domains of width \(\approx 3\sigma_{0}\). The strong dipolar interactions in this case stabilizes a biaxial phase instead of a columnar phase. The characteristics are again found qualitatively similar to that exhibited by (N = 4000,\(\mu^{*}=0.90\)) system studied using the spherical cutoff approach. The plots of the structural distribution functions are given in Fig. \ref{fig:p98}. The flatness in the radial distribution function \(g(r^{*})\) at \(T^{*}=11\) confirms the absence of any long-range positional order in the striped nematic phase. Figs. \ref{fig:p98b} and \ref{fig:p98c} show that the columnar and perpendicular distribution functions remains essentially flat for the nematic phase at \(T^{*}=11\). Considerable structures in \(g_{c}(r_{\shortparallel}^{*})\) and \(g(r_{\perp}^{*})\) at \(T^{*}=10.5\) indicate the formation of the higher ordered biaxial phase.  We have measured the biaxial order parameter \( \langle R_{2,2}^{2}\rangle \), the value which is \(\approx\) 0.95 as measured in the striped biaxial phase at \(T^{*}=10.5\). However, \( \langle R_{2,2}^{2}\rangle \) remain \(\approx\) 0 in the striped nematic phase. The columnar and transverse distribution functions at \(T^{*}=10.5\) are found quite different from those obtained for the columnar phases. From the snapshot of the biaxial phase in Fig. \ref{fig:p21d}, we can understand that the structure is quite similar to the biaxial phase obtained using the spherical cutoff approach. It can be seen that the dipoles are arranged such that the dipolar separation vectors pointing from one dipole on a disc to the other get oriented mostly in same direction. This ordering can be understood as the effect of strong pair interaction possible in this condition between two terminal dipoles of neighboring ellipsoids if two parallel dipoles are placed above each other. At \(T^{*}=10.5\), \(g_{c}(r_{\shortparallel}^{*})\) indicates the intercalated arrangement of the ellipsoids. \\

 The formation of a biaxial phase via dipolar interactions supports earlier simulation works on dipolar anisotropic particles which showed that dipolar interactions can introduce a biaxial orientational order even in the absence of a shape biaxiality of the constituent molecules\cite{b282,b283}. The global ferroelectric order parameter\( \langle P_{1} \rangle\) remained approximately equal to zero indicating zero net polarization for all the phases observed for \(\mu^{*}=0.90\). We have observed that the system developed a small cavity at this dipole strength which might be a result of the fixed shape constraint of the cubic box used by us as reported earlier in studies of discotic liquid crystals in \cite{b331}.\\

Therefore, the simulations of smaller systems performed using the ES technique, generated results qualitatively similar to that obtained from the simulations of larger systems using the spherical cutoff technique. Usually it is considered that the macroscopic properties of a system of dipolar particles converge more rapidly with systemsize in case of ES method (\(\epsilon_{s}=\infty\)) than in case of RF method. Here, the results also indicate that the macroscopic properties of a system of dipolar particles exhibiting long range ferroelectric order also converge more rapidly in case of the ES method (\(\epsilon_{s}=1\)) than in case of the simpler spherical cutoff method. A study on the system size effects using ES technique, although of undeniable interest, has not been attempted here because of the substantial computing cost required. However, we think that the domain size would increase with system size in the ES method similar to behavior observed in spherical cutoff approach. A similar view was also discussed for ferroelectric systems of dipolar spheres in \cite{b21}.\\
\begin{figure}[!]
\centering
\includegraphics[scale=1.0]{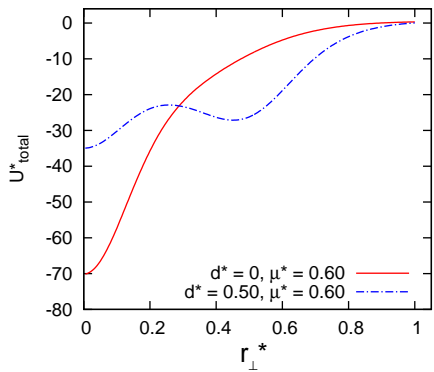}

\caption{\label{fig:p153}(color online). The variation of the pair interaction energy \(U^{*}_{total}\) of two polar ellipsoids having parallel dipolar separation vectors as one polar ellipsoid is slid over another along dipolar separation vector with their dipoles oriented along their symmetry axes and keeping the component of the inter particle vector parallel to the symmetry axes of the particles at a fixed value of \(\sigma_{e}\). \(r_{\perp}^{*}\) is the component of the inter particle vector perpendicular to the symmetry axes of the particles. The different curves are for different combinations of \(d{*}\) and \(\mu^{*}\) as described inside the figure.}
\end{figure}

Let us now discuss the behavior of the pair interactions. The pair interaction energy between two dipolar ellipsoids is plotted in fig. \ref{fig:p153} as a function of the component of the pair separation vector \(\mathbf r\) perpendicular to the particle symmetry axes  \(r_{\perp}^{*}\) as one particle is slid over the other maintaining a fixed separation (\( r_{\shortparallel}=\sigma_{e}\)) along the symmetry axis and having all dipoles oriented in same direction. Note that when \(r_{\perp}^{*}\) is zero one particle completely covers the other. Curves are shown for \(d^{*}=0 \mbox{ and }d^{*}=0.50\). Here we have discussed the case where the ellipsoid is sliding along the direction of molecular x axis, i.e. along the dipolar separation vector \(\mathbf d^{*}(=d^{*}\hat{x})\) of the ellipsoids. This particular case is important for the biaxial phase behavior found for \(\mu^{*}\)=0.90. It is evident from Fig. \ref{fig:p153} that the dipolar separation \(d^{*}\) has a dramatic role on the pair potential. For \(d^{*}=0\), the strong head-to-tail dipolar interaction results in a sharp minima at \(r_{\perp}^{*}\)= 0. It is the sharpness of this potential which leads to strong stabilization of the columnar phase for central dipolar disk shaped particles studied in \cite{b11,b765}. In the columnar phases of central dipolar disks, the individual polarized columns do not interact strongly enough and entropy ensures that equal numbers will be polarized in opposite directions \cite{b11,b765}. In the present model, the pair interaction looses its sharp focus when the dipoles are more separated and generates a weaker and slowly varying attraction which helps in generating the ferroelectric nematic order for \(d^{*}\)= 0.50. It can be seen that for separation \(d^{*}\)= 0.50, a second well is constructed with its minimum at \(r_{\perp}^{*}\)=0.50. The second well simply results from the interaction between two parallel dipoles when they are positioned above each other. The strengthening of this second well favors strong stabilization of the ferroelectric biaxial phase at higher dipole strengths. In the uniaxial striped nematic and striped columnar phases, the dipolar separation vectors of two neighboring ellipsoids are usually randomly oriented with respect to each other. In such cases, the pair interaction do not generate a second well as found for parallel orientation of the dipolar separation vectors of the two ellipsoids. Here, the formation of the ferroelectric columnar domains for \(\mu^{*}=0.40,0.60 \mbox{ and }0.70\) indicate that the column-column interactions are stronger than that in case of central dipolar disks.\\ 


\section{\label{sec:level5}Conclusions }

In this work, we have systematically investigated the influences of the depolarizing boundary conditions upon the existence of a class of novel ferroelectric phases of dipolar origin in the systems of disklike anisotropic particles using both the spherical cutoff and the Ewald Summation techniques. 
The systems resulted in the formation of periodic slablike ferroelectric fluid domains arranged in an antiferroelectric fashion. 
We found existence of the long range ferroelectric nematic, ferroelectric columnar and ferroelectric biaxial orderings within the domain regions at different state points and dipole strengths. It is found that the minimum value of the dipole strength required to generate a ferroelectric order is 
\(\mu^{*} = 0.40\). The periodicity and the size of the slablike domains, for a fixed systemsize, remained unchanged with respect to a variation of temperature or dipole strength studied here. It is explicitly shown that the size of the ferroelectric domains grow with the system size as an effect of considering increasing range of dipolar interactions. These results provide great support to the novel understanding that the dipolar interactions are indeed sufficient to produce a long range ferroelectric order in discotic fluids. Our results also supports the view expressed in \cite{b21} that for sufficiently large samples the local orientational order in the ferroelectric domains would be quite similar to that obtained with \(\epsilon_{s} = \infty\). Our study should be helpful in analyzing and understanding the structural properties of novel anisotropic fluids where dipolar forces would play a dominant role in generating global ferroelectric order. 



\section{ ACKNOWLEDGEMENT }

T.K. Bose gratefully acknowledges the support of CSIR, India for providing Senior Research Fellowship via sanction no. 09/028(0794)/2010-EMR-I. This work was supported by the UGC-UPE scheme of the University of Calcutta and the FIST scheme of DST, India.

\FloatBarrier

\bibliographystyle{plainnat}.

\bibliography{PRE}

\end{document}